\pdfoutput=1
\documentclass[oldversion]{aa}
\usepackage{graphicx}
\usepackage{rotating}
\usepackage{txfonts}
\usepackage{epsf}
\usepackage{natbib}
\usepackage{hyperref}

\def\new#1{{\bf \small #1}}

\def\oii{{\sc{[Oii]}}}
\def\sigmamap{$\sigma$-map }

\def\zzz{$z\!=\!0.4\,\textrm{--}\,0.75$}

\begin{document}
\title{IMAGES\thanks{Intermediate MAss Galaxy Evolution Sequence, ESO programs 174.B-0328(A), 174.B-0328(A), 174.B-0328(E)}
 I. Strong evolution of galaxy kinematics since $z\!=\!1$
 }
\author{  
       Y. Yang\inst{1}                     %Yanbin YANG <yanbin.yang@obspm.fr>                           
  \and H. Flores\inst{1}                   %Hector Flores <hector.flores@obspm.fr>                               
  \and F. Hammer\inst{1}                   %Francois Hammer <francois.hammer@obspm.fr>                   
  \and B. Neichel\inst{1}                  %Benoit Neichel <benoit.neichel@obspm.fr>                     
  \and M. Puech\inst{2}                    %PUECH <mpuech@eso.org>                                               
  \and N. Nesvadba\inst{1}                 %Nicole Nesvadba <nicole.nesvadba@obspm.fr>                   
  \and A. Rawat\inst{1,3}                  %Abhishek Rawat <rawat@iucaa.ernet.in>                                
  \and C. Cesarsky\inst{2}                 %Catherine Cesarsky <ccesarsk@eso.org>                                
  \and M. Lehnert\inst{1}                  %Matt Lehnert <matthew.lehnert@obspm.fr>                              
  \and L. Pozzetti\inst{4}                 %lucia.pozzetti@oabo.inaf.it                                  
  \and I. Fuentes-Carrera\inst{1}          %isaura.fuentes@obspm.fr                                              
  \and P. Amram\inst{5}                    %Philippe.Amram@oamp.fr                                               
  \and C. Balkowski\inst{1}                %Chantal Balkowski <Chantal.Balkowski@obspm.fr>                       
  \and H. Dannerbauer\inst{6}              %Helmut Dannerbauer <dannerb@mpia-hd.mpg.de>                  
  \and S. di Serego Alighieri\inst{7}      %Sperello di Serego Alighieri <sperello@arcetri.astro.it>     
  \and B. Guiderdoni\inst{8}             %Bruno Guiderdoni <bguider@obs.univ-lyon1.fr>                         
  \and A. Kembhavi\inst{3}                 %Ajit Kembhavi <akk@iucaa.ernet.in>                           
  \and Y. C. Liang\inst{9}                %Yanchun Liang <ycliang@bao.ac.cn>                            
  \and G. {\"O}stlin\inst{10}              %Goeran Ostlin <ostlin@astro.su.se>                           
  \and C. D. Ravikumar\inst{11}            %--? cdravi@gmail.com, Chazhiyat.Ravikumar@obspm.fr                       
  \and D. Vergani\inst{12}                 %--? daniela@mi.iasf.cnr.it
  \and J. Vernet\inst{2}                   %Joel Vernet <jvernet@eso.org>
  \and H. Wozniak\inst{8}                  %herve.wozniak@obs.univ-lyon1.fr                                  
}                                          
%%Y. Yang, H. Flores, F. Hammer, B. Neichel, M. Puech, N. Nesvadba, A. Rawat, C. Cesarsky, M. Lehnert, L. Pozzetti, I. Fuentes-Carrera, P. Amram, C. Balkowski, H. Dannerbauer, S. di Serego Alighieri, B. Guiderdoni,A. Kembhavi, Y. C. Liang, G. {\"O}stlin, C. D. Ravikumar, D. Vergani, J. Vernet, H. Wozniak
\offprints{yanbin.yang@obspm.fr}
\authorrunning{Y. Yang et al.}
\titlerunning{IMAGES I. Observations}
\institute{
  %1
GEPI, Observatoire de Paris, CNRS, University Paris Diderot; 5 Place Jules Janssen, Meudon, France 
\and %2
ESO, Karl-Schwarzschild-Strasse 2, D-85748 Garching bei M\"unchen, Germany
\and %3
Inter-University Centre for Astronomy and Astrophysics, Post Bag 4, Ganeshkhind, Pune 411007, India 
\and %4
 INAF - Osservatorio Astronomico di Bologna, via Ranzani 1, 40127 Bologna, Italy
\and %5
Laboratoire d'Astrophysique de Marseille, Observatoire Astronomique de 
Marseille-Provence, 2 Place Le Verrier, 13248 Marseille, France
\and %6
MPIA, K{\"o}nigstuhl 17, D-69117 Heidelberg, Germany
\and %7
INAF, Osservatorio Astrofisico di Arcetri, Largo Enrico Fermi 5, I-50125, Florence, Italy
\and %8
Centre de Recherche Astronomique de Lyon, 9 Avenue Charles André, 69561 Saint-Genis-Laval 
Cedex, France 
%\and %9,
%Institut d'Astrophysique du CNRS, 98 bis Boulevard Arago, F-75014 Paris, France
\and %10
National Astronomical Observatories, Chinese Academy of Sciences, 20A Datun Road, Chaoyang District, Beijing 100012, PR China 
\and %11 
Stockholm Observatory, AlbaNova University Center, Stockholms Center for Physics, Astronomy and Biotechnology, Roslagstullsbacken 21, 10691 Stockholm, Sweden 
\and %12 
Department of Physics, University of Calicut, Kerala 673635, India
\and %13
IASF-INAF - via Bassini 15, I-20133, Milano, Italy
}
\date{Received ...... ; accepted ...... }

\abstract{{}{Nearly half the stellar mass of present-day spiral
galaxies has formed since $z\!=\!1$, and galaxy kinematics is an ideal
tool to identify the underlying mechanisms responsible for the galaxy
mass assembly since that epoch.} {Here, we present the first results
of the ESO large program, ``IMAGES'', which aims at obtaining robust
measurements of the kinematics of distant galaxies using the multi-IFU
mode of GIRAFFE on the VLT. 3D spectroscopy is essential to robustly
measure the often distorted kinematics of distant galaxies (e.g.,
Flores et al.\ 2006).  We derive the velocity fields and $\sigma$-maps
of 36 galaxies at $0.4\!<\!z\!<\!0.75$ from the kinematics of the
\oii\ emission line doublet, and generate a robust technique to
identify the nature of the velocity fields based on the pixels of the
highest signal-to-noise ratios (S/N). } {Combining these observations with
those of Flores et al., we have gathered a unique sample of 63
velocity fields of emission line galaxies ($W_{0}(\mbox{{\sc
[Oii]}})\ge15$\,\AA) at $z\!=\!0.4$\,--\,0.75, which are a
representative subsample of the population of $M_{\rm
stellar}\!\ge\!1.5\!\times\!10^{10}M_{\sun}$ emission line galaxies in
this redshift range, and are largely unaffected by cosmic
variance. Taking into account all galaxies -with or without emission
lines- in that redshift range, we find that at least 41$\,\pm\,$7\% of
them have anomalous kinematics, i.e., they are not dynamically
relaxed. This includes 26$\,\pm\,$7\% of distant galaxies with complex
kinematics, i.e., they are not simply pressure or rotationally
supported.}  {Our result implies that galaxy kinematics are among the
most rapidly evolving properties, because locally, only a few percent
of the galaxies in this mass range have complex kinematics. It is
well-established that galaxies undergoing a merger have complex
large-scale motions and thus are likely responsible for the strong
evolution of the galaxy kinematics that we observe.}

}

\maketitle

%\tableofcontents

\section{Introduction} 
The resolved 3D kinematics of distant galaxies are a powerful tracer of
the major processes governing star-formation and galaxy evolution in
the early universe such as merging, accretion, and hydrodynamic
feedback related to star-formation and active galactic nucleus (e.g., Barnes \& Hernquist
1996; Barton et al.\ 2000; Dressler 2004). Thus, robustly measuring
the internal kinematics of galaxies in the distant universe plays a
crucial role for our growing understanding of how galaxies formed and
evolved.

Over the last decade, great efforts have been made to study the
properties of galaxies in the distant universe (at $z\!\sim\!1$),
revealing a strong evolution with cosmic time. For instance, the
cosmic star formation rate (SFR) has declined by a factor $\sim\!10$
from $z\!\sim\!1$ to the present (Lilly et al.\ 1996; Madau et al.\
1996; Hammer et al.\ 1997; Cowie et al.\ 1999; Flores et al.\ 1999).
Such a strong evolution of cosmic SFR is consolidated by
subsequent works, e.g., Haarsma et al.\ (2000), Wilson et al.\
(2002). Although the conclusions are made from different data, they
are consistent within a factor of 3 (Hopkins 2004).  Heavens et
al.\ (2004) suggest that the cosmic SFR may have reached its peak as
late as $z\!\sim\!0.6$. Overall, about half of the stellar mass in
intermediate-mass galaxies was formed since $z\!=\!1$, mostly in
luminous infrared galaxies (Hammer et al.\ 2005).

%What were the physical processes driving star-formation at these redshifts?  
Galaxy interactions and merging may be mechanisms that played a
significantly larger role for star-formation in the distant universe
than today.  Le F{\`e}vre et al.\ (2000) found that the merger rate in
the distant universe was about a factor of 10 times higher than at low
redshift (see also Conselice et al.\ 2003; Bell et al.\ 2006;
Lotz et al.\ 2006).  The high merger rate detected in the distant
universe raises a challenge to the standard scenario of disk formation
(Hammer et al.\ 2007).  If major mergers generate the ellipticals
inevitably, we would find a large fraction of elliptical galaxies rather
than about $\sim$70\% of spiral galaxies among the intermediate-mass
galaxies in the local universe.  Similarly, the fraction of luminous
compact blue galaxies (LCBGs) increases with redshift by about an
order of magnitude out to $z\!\sim\!1$ (Werk et al.\ 2004; Rawat et
al.\ 2007).  LCBGs may be the progenitors of local spheroidal or
irregular galaxies at low redshift (e.g., Koo et al.\ 1995; Guzman
1999), or of the bulges of massive spirals (Hammer et al.\ 2001;
Noeske et al.\ 2006).  3D spectroscopy of the internal kinematics of
LCBGs suggests that they are likely merger remnants ({\"O}stlin et
al. 2001; Puech et al.\ 2006a).

Strong evolution as a function of cosmic time has also been claimed
for the Tully-Fisher relationship (TFR, Tully \& Fisher 1977;
Giovanelli et al.\ 1997), which relates the luminosity and the
rotation velocity of disk galaxies. Out to $z\!\sim\!1$, the B-band
TFR has been found to have evolved by $\sim\!0.2$\,--\,$2$ mag (e.g.,
Portinari \& Sommer-Larsen 2007 and references therein). This
brightening of the B-band TFR can be explained by the enhanced
star-formation rates at higher redshifts (Ferreras \& Silk 2001;
Ferreras et al.\ 2004), but is still a matter of debate. Conselice et
al.\ (2005) do not find significant evolution in either the stellar
mass or K-band TFR's slope or zero point. However, the most striking
evolution of the TFR is provided by its large scatter at high
redshifts (Conselice et al.\ 2005), which may be related to the
disturbed kinematics of distant galaxies (e.g., Kannappan \& Barton
2004).

The rapid time decrease of cosmic SFR, the role of merging in the early evolution
of galaxies, and the possible evolution of the TFR 
are only examples of why measuring the kinematics of distant galaxies
precisely and robustly is a {\it sine qua non} for studying galaxy
evolution. However, this is often beyond what can be achieved with
classical long-slit spectroscopy. The morphologies and kinematics of
distant galaxies are often complex, and their small sizes make it very
difficult to precisely position and align the slit. Both limitations
can be overcome with integral-field spectroscopy, although the method
is relatively complex and time-consuming.

Flores et al.\ (2006) presented the first study of a statistically
meaningful sample of 35 intermediate-mass galaxies at z=0.4-0.7, using
the integral-field 
multi-object spectrograph GIRAFFE on the ESO-VLT. They defined a
classification scheme to distinguish between rotation and kinematic
perturbances, which may stem from interactions and mergers, from the
3D kinematics and high-resolution HST imaging.
Intriguingly, they find that the large scatter of distant TFR shown in
previous studies is due to non-relaxed systems while the pure
rotational disks exhibit a TFR that is similarly tight as that of
local spirals. Here, we present another sample of 36 galaxies with very
similar selection criteria, to enlarge the total sample size and put
the conclusions on statistically more robust grounds. This is the
first of a series of publications related to the ESO-VLT large program
IMAGES, which aims at studying the evolutionary sequence of galaxies
over the last 8 Gyrs (see Ravikumar et al.\ 2007 for more details).

The paper is organized as follows. In Sect.~2 we describe the
observations and the sample selection. The methodology to describe and
classify the distant galaxy kinematics is shown in Sect.~3, as
well as a detailed description of the 36 observed velocity
fields (VFs). Sections 4 and 5 include the discussion and the conclusion. In
this paper, we adopt the Concordance cosmological parameters of
$H_0\!=\!70$ km s$^{-1}$ Mpc$^{-1}$, $\Omega_M\!=\!0.3$ and
$\Omega_\Lambda\!=\!0.7$.

{\scriptsize
\begin{table}[!t]
\caption{Journal of observations.}
\begin{center}
\begin{tabular}{lccr}\hline
Run ID & Setup & Exposure (hr) \\\hline
174.B-0328(A) &   L04 & 10   \\
073.A-0209(A) &   L05 & 4.5  \\
174.B-0328(A) &   L05 & 10.6 \\
174.B-0328(E) &   L05 & 10.4 \\\hline
\end{tabular}
\end{center}
\label{tbobservation}
\end{table}
}

\section{Data}
\subsection{Observations}
We used the FLAMES-GIRAFFE multi-object integral-field spectrograph on
the ESO-VLT in the multi-IFU mode to measure the velocity and
dispersion fields of a statistically meaningful sample of galaxies at
redshifts \zzz\ from their {\sc [Oii]}$\lambda$3726,3729
emission. Each integral-field unit (IFU) of GIRAFFE consists in 20
micro-lenses with 0.52\arcsec\ spatial sampling, resulting in a
$2\arcsec\!\times\!3$\arcsec\ field of view per IFU. We used the LR04
and LR05 set-ups, which correspond to spectral resolutions of 0.55\,\AA\
(30 km\,s$^{-1}$) and 0.45\,\AA\ (22 km\,s$^{-1}$), respectively.

Observations were carried out as part of the IMAGES large program,
complemented by guaranteed time observations (programs 174.B-0328(A),
073.A-0209(A), 174.B-0328(A), 174.B-0328(E), see also
Table~\ref{tbobservation}). The total observing time was 5 nights,
with integration times ranging from 4.5 to 15 hrs for individual
targets. The seeing ranged from 0.4\arcsec to 1\arcsec, with a median
value of 0.8\arcsec. Data reduction and the construction of the final
data cubes are described in detail in Flores et al.\ (2006).

\subsection{Sample selection}
Our targets are a subset of the Chandra Deep Field South, with
redshifts $z\!\sim\!0.4$--\,$0.75$, $I_{\rm AB}\!\le\!23.5$ and detected
{\sc [Oii]}$\lambda$3726,3729 emission lines ($W_{0}(\mbox{{\sc
[Oii]}})\ge15$\,\AA, Ravikumar et al.\ 2007). Our goal is to investigate
a sample of intermediate mass galaxies (see Hammer et al.\ 2005), 
therefore we required $J$-band absolute magnitudes brighter than
$M_J{\rm{(AB)}}\! =\! -20.3$. Such a limit corresponds approximately
to a stellar mass of $M_{\rm stellar} \ge 1.5\times 10^{10}M_{\sun}$
when converting the $J$-band luminosity using the prescription
discussed in Bell et al. (2003; see also Hammer et al.\ 2005).
Ravikumar et al. (2007) has convincingly shown that within the
redshift range of 0.4--\,0.75, $I_{AB}\le 23.5$ galaxies include
almost all intermediate mass galaxies (e.g., at least 95\% of $M_J{\rm{(AB)}}$ $\le$ -20.3 galaxies,
see their Sect. 3.4). Given all the above, our sample comprises a 
total of 46 targets. The relatively small number of
suitable galaxies and small bandpass of the GIRAFFE set-ups make it
difficult to fill all 15 IFUs with galaxies with spectroscopic
redshifts. We therefore used empty ``bonus'' IFUs to observe galaxies
for which only photometric redshifts were known, but could not detect
any because of the large uncertainties of photometric redshifts.

Moreover, we rejected 2 galaxies with spurious features that were identified as \oii\
emission lines in their spectra (J033221.42-274231.2, J033241.08-274853.0), 4
targets due to faint line emission (i.e., $W_{0}(\mbox{\sc
[Oii]})\!<\!15$\,\AA: J033211.41-274650.0, J033226.00-274150.6,
J033232.13-275105.5, J033254.50-274703.6), and one due to the CCD
defects (J033213.85-274248.9). Another 2 galaxies
(J033212.36-274835.6, J033236.72-274406.4) were rejected by our
minimum quality criterion: at least 4 GIRAFFE spatial pixels with
spectral S/N\,$>$\,4. The galaxy
J033229.71-274507.2 was also rejected because emission was detected in
only 4 GIRAFFE pixels, which is not sufficient to classify its
kinematics. Finally, we obtained a sample of 36 well resolved galaxies
of intermediate stellar mass with good S/N values.

\subsection{Representativeness/completeness of the sample}
\label{secrepsample}
\begin{figure}[!t]
\begin{center}
\includegraphics[width=7cm]{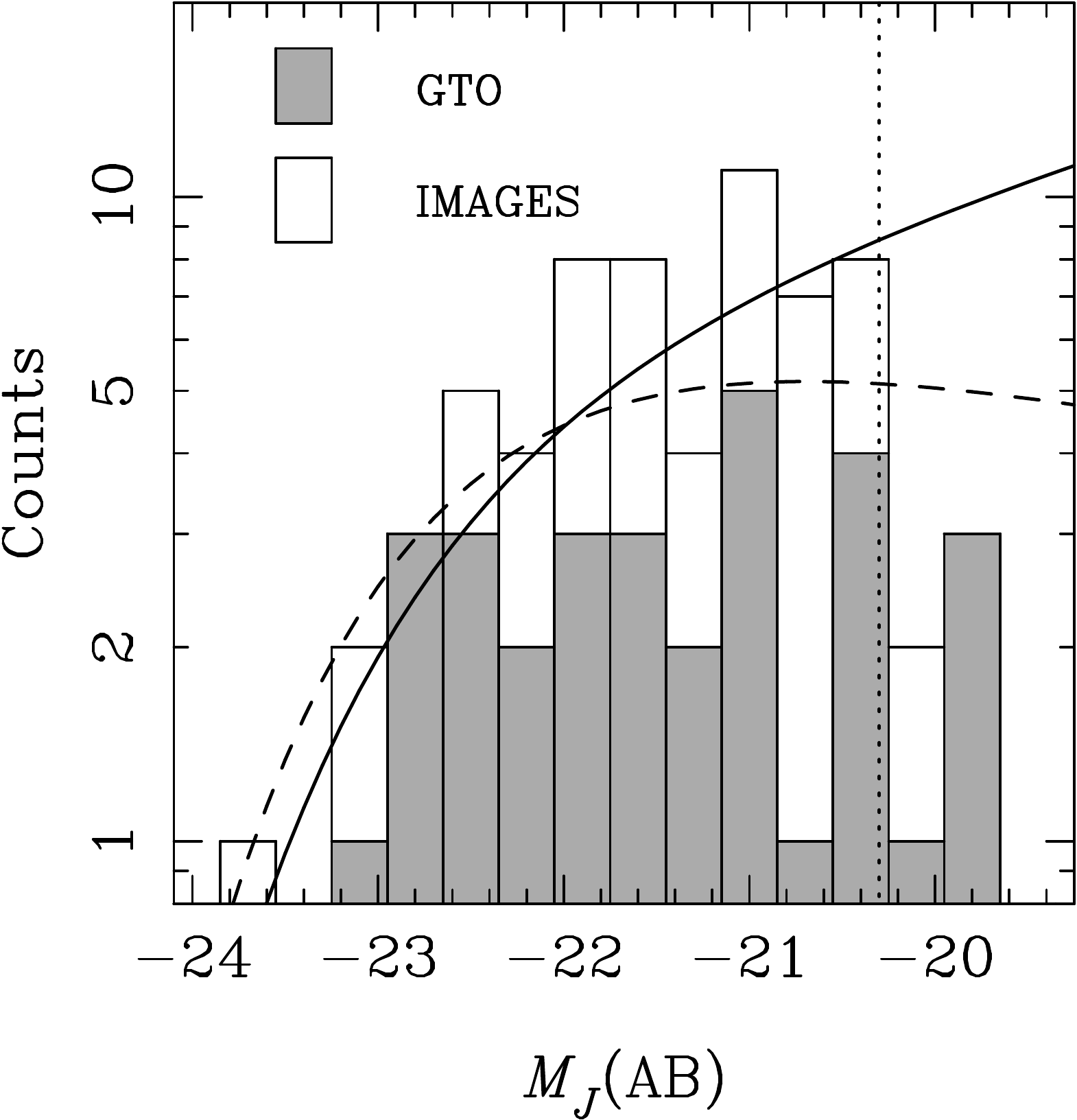}
\caption{Number counts (in logarithmic scale) of selected galaxies
versus AB absolute magnitude in $J$-band. The GTO sample refers to
Flores et al.\ (2006); the IMAGES sample refers to this paper. The vertical
dotted line indicates the limit of the IMAGES program. Two luminosity
functions derived from Pozzetti et al.\ (2003) are shown (full line:
$z\!=\!0.5$; dashed line: $z\,=\,1$). The galaxies of our sample have
redshifts ranging from $z\!=\!0.4$ to $z\!=\!0.75$. This implies that
the combined sample of 63 galaxies with $M_J{\rm{(AB)}} \le -20.3$ is
representative of galaxies with stellar masses larger than $1.5\times
10^{10}M_{\sun}$ at $z\sim0.6$.}
\end{center}
\label{fighist}
\end{figure}

Figure \ref{fighist} shows the distribution of the $J$-band absolute
magnitudes (also listed in Table~\ref{tbobjects}) for the sample
studied in this paper, combined with the Flores et al.\ (2006) sample
of 35 galaxies. Both samples can be merged because the selection of
Flores et al.\ is very similar to that of this paper and because both
studies used essentially the same instrumental set-ups. Applying our
criteria of $M_J{\rm{(AB)}}\!\le\!-20.3$ and $W_{0}(\mbox{\sc
[Oii]})\!\ge\!15$\,\AA, we are left with 63 galaxies with data of
appropriate quality to carry out our analysis. We compared the
luminosity distribution of the sample to the luminosity function at
redshift of 0.5 and 1 (Fig.~\ref{fighist}).  Kolmogorov-Smirnov tests
support that our sample follows the luminosity function in the \zzz\
range at $>99.9\%$ confidence level.  Furthermore, the combined two samples
include galaxies from 4 different fields, namely the CDFS, HDFS,
CFRS03h and CFRS22h. It is then unlikely that our
conclusions are strongly affected by statistical effects, possibly
related to large scale structures (see Sect.~\ref{secdis} for more
analysis, also Ravikumar et al.\ 2007). Our sample is a completely
representative sub-sample of $M_J{\rm{(AB)}}\!\le\!-20.3$
emission line selected galaxies at \zzz. To our
knowledge, it is the only such existing sample of distant galaxies
with measured 3D kinematics.

{\scriptsize
%\begin{sidewaystable*}
\begin{table*}[!t]
\caption{Properties of the 39 galaxies of the IMAGES sample.}
\begin{center}
\begin{tabular}{cccccr@{}lcr@{}lr@{}l}\hline\hline
      GOODS ID      &      RA \& DEC (J2000.0)   &     $z^{\rm{a}}$    &  $M_{B}$$^{\rm{b}}$&  $M_J$$^{\rm{b}}$&  &    $i^{\rm{c,e}}$ &  C$^{\rm{d}}$ &  $\Delta$&$r$$^{\rm{e}}$   & $\epsilon$$^{\rm{e}}$ \\ \hline 
J033212.39-274353.6 &  03:32:12.387 $-$27:43:53.59 &  0.42130 & $-$19.55 & $-$21.58  &  90.0  &(5.0)  &  RD &0.08 & (0.32) & 0.13 & (0.03)   \\           
J033219.68-275023.6 &  03:32:19.678 $-$27:50:23.57 &  0.55955 & $-$20.88 & $-$22.37  &  58.3  &(6.7)  &  RD &0.11 & (0.46) & 0.06 & (0.01)   \\           
J033230.78-275455.0 &  03:32:30.780 $-$27:54:54.99 &  0.68573 & $-$20.51 & $-$21.90  &  66.1  &(4.3)  &  RD &0.11 & (0.31) & 0.12 & (0.02)   \\           
J033231.58-274121.6 &  03:32:31.575 $-$27:41:21.63 &  0.70411 & $-$20.16 & $-$20.69  &  42.2  &(9.6)  &  RD &0.11 & (0.57) & 0.01 & ($<$.01) \\           
J033234.04-275009.7 &  03:32:34.037 $-$27:50:09.69 &  0.70162 & $-$19.88 & $-$20.61  &  59.3  &(3.4)  &  RD &0.09 & (0.21) & 0.17 & (0.02)   \\           
J033237.54-274838.9 &  03:32:37.538 $-$27:48:38.94 &  0.66377 & $-$21.29 & $-$22.07  &  30.7  &(12.)  &  RD &0.11 & (0.35) & 0.04 & (0.01)   \\           
J033238.60-274631.4 &  03:32:38.595 $-$27:46:31.37 &  0.62066 & $-$20.03 & $-$21.54  &  60.2  &(4.3)  &  RD &0.16 & (0.34) & 0.08 & (0.02)   \\           
J033241.88-274853.9 &  03:32:41.883 $-$27:48:53.86 &  0.66702 & $-$20.32 & $-$21.00  &  66.5  &(6.2)  &  RD &0.03 & (0.50) & 0.08 & (0.01)   \\           
J033245.11-274724.0 &  03:32:45.108 $-$27:47:24.00 &  0.43462 & $-$20.13 & $-$22.06  &  43.3  &(6.2)  &  RD &0.14 & (0.18) & 0.09 & (0.02)   \\ \hline    
J033210.25-274819.5 &  03:32:10.250 $-$27:48:19.49 &  0.60874 & $-$19.76 & $-$20.93  &  68.5  &(4.5)  &  PR &0.83 & (0.19) & 0.40 & (0.07)   \\           
J033214.97-275005.5 &  03:32:14.971 $-$27:50:05.45 &  0.66652 & $-$21.50 & $-$22.53  &  21.9  &(8.2)  &  PR &1.32 & (0.27) & 0.65 & (0.12)   \\           
J033219.61-274831.0 &  03:32:19.606 $-$27:48:30.97 &  0.66992 & $-$20.36 & $-$20.99  &  49.1  &(7.0)  &  PR &0.40 & (0.35) & 0.18 & (0.01)   \\           
J033226.23-274222.8 &  03:32:26.229 $-$27:42:22.81 &  0.66713 & $-$20.71 & $-$22.01  &  76.4  &(3.1)  &  PR &0.72 & (0.42) & 0.40 & (0.09)   \\           
J033232.96-274106.8 &  03:32:32.959 $-$27:41:06.78 &  0.46811 & $-$19.50 & $-$20.45  &  16.0  &(3.8)  &  PR &0.64 & (0.37) & 1.15 & (0.14)   \\           
J033233.90-274237.9 &  03:32:33.897 $-$27:42:37.93 &  0.61801 & $-$20.99 & $-$21.91  &  17.2  &(10.)  &  PR &0.07 & (0.26) & 1.35 & (0.17)   \\           
J033239.04-274132.4 &  03:32:39.044 $-$27:41:32.43 &  0.73186 & $-$20.46 & $-$20.75  &  43.2  &(10.)  &  PR &0.44 & (0.25) & 0.24 & (0.01)   \\           
J033243.62-275232.6 &  03:32:43.623 $-$27:52:32.63 &  0.67823 & $-$19.27 & $-$20.03  &  70.8  &(1.6)  &  PR &0.15 & (0.63) & 1.54 & (0.25)   \\           
J033248.28-275028.9 &  03:32:48.281 $-$27:50:28.88 &  0.44464 & $-$19.35 & $-$20.47  &  80.8  &(3.3)  &  PR &0.34 & (0.82) & 0.40 & (0.04)   \\           
J033249.53-274630.0 &  03:32:49.525 $-$27:46:29.98 &  0.52212 & $-$20.05 & $-$21.09  &  45.6  &(2.2)  &  PR &1.10 & (0.58) & 0.38 & (0.03)   \\           
J033250.53-274800.7 &  03:32:50.534 $-$27:48:00.67 &  0.73604 & $-$19.99 & $-$20.50  &  62.3  &(3.9)  &  PR &0.22 & (0.62) & 0.30 & (0.04)   \\ \hline    
J033210.76-274234.6 &  03:32:10.761 $-$27:42:34.58 &  0.41686 & $-$21.78 & $-$23.70  &  26.0  &(7.4)  &  CK &0.74 & (0.54) & 0.47 & (0.09)   \\           
J033213.06-274204.8 &  03:32:13.061 $-$27:42:04.81 &  0.42150 & $-$19.53 & $-$20.67  &  78.7  &(2.8)  &  CK &0.68 & (0.30) & 0.02 & ($<$.01) \\           
J033217.62-274257.4 &  03:32:17.620 $-$27:42:57.44 &  0.64565 & $-$19.78 & $-$21.23  &  46.3  &(8.4)  &  CK &1.07 & (0.21) & 0.04 & (0.01)   \\           
J033219.32-274514.0 &  03:32:19.317 $-$27:45:14.04 &  0.72411 & $-$20.31 & $-$21.24  &  72.1  &(4.5)  &  CK &0.15 & (0.17) & 0.66 & (0.12)   \\           
J033220.48-275143.9 &  03:32:20.484 $-$27:51:43.93 &  0.67780 & $-$19.93 & $-$20.72  &  63.4  &(3.6)  &  CK &1.05 & (0.30) & 2.94 & (0.42)   \\           
J033224.60-274428.1 &  03:32:24.601 $-$27:44:28.12 &  0.53680 & $-$19.58 & $-$20.44  &  64.8  &(1.2)  &  CK &0.99 & (0.48) & 0.42 & (0.04)   \\           
J033225.26-274524.0 &  03:32:25.260 $-$27:45:23.97 &  0.66479 & $-$20.90 & $-$21.63  &  60.5  &(6.0)  &  CK &0.87 & (0.19) & 1.00 & (0.09)   \\           
J033227.07-274404.7 &  03:32:27.074 $-$27:44:04.66 &  0.73814 & $-$20.34 & $-$21.04  &  84.2  &(1.7)  &  CK &0.69 & (0.14) & 0.91 & (0.13)   \\           
J033228.48-274826.6 &  03:32:28.477 $-$27:48:26.55 &  0.66857 & $-$20.10 & $-$21.74  &  22.4  &(5.0)  &  CK &0.67 & (0.24) & 1.35 & (0.26)   \\           
J033230.43-275304.0 &  03:32:30.429 $-$27:53:04.02 &  0.64533 & $-$19.96 & $-$21.71  &  70.3  &(2.4)  &  CK &1.02 & (0.24) & 0.44 & (0.09)   \\           
J033230.57-274518.2 &  03:32:30.569 $-$27:45:18.24 &  0.67988 & $-$21.93 & $-$22.95  &  34.5  &(9.4)  &  CK &1.04 & (0.52) & 0.16 & (0.02)   \\           
J033234.12-273953.5 &  03:32:34.120 $-$27:39:53.53 &  0.62734 & $-$23.08 & \ \ 99.99 &  32.1  &(4.5)  &  CK &0.12 & (0.18) & 0.53 & (0.08)   \\           
J033239.72-275154.7 &  03:32:39.719 $-$27:51:54.68 &  0.41510 & $-$20.10 & $-$21.04  &  35.4  &(1.3)  &  CK &1.55 & (0.45) & 5.15 & (0.26)   \\           
J033240.04-274418.6 &  03:32:40.040 $-$27:44:18.63 &  0.52201 & $-$20.55 & $-$22.04  &  15.6  &(11.)  &  CK &0.37 & (0.22) & 0.36 & (0.08)   \\           
J033244.20-274733.5 &  03:32:44.199 $-$27:47:33.48 &  0.73605 & $-$21.08 & $-$21.86  &  39.0  &(5.8)  &  CK &0.64 & (0.15) & 0.62 & (0.03)   \\           
J033250.24-274538.9 &  03:32:50.239 $-$27:45:38.92 &  0.73099 & $-$19.79 & $-$20.70  &  31.3  &(8.8)  &  CK &1.37 & (0.38) & 0.59 & (0.08)   \\ \hline    
J033212.36-274835.6 &  03:32:12.360 $-$27:48:35.64 &  0.56210 & $-$20.13 & $-$21.16  &  65.8  & (1.4) &  UC &  &--  &   & --  \\   
J033229.71-274507.2 &  03:32:29.707 $-$27:45:07.20 &  0.73170 & $-$20.05 & $-$20.88  &  42.3  & (6.2) &  UC &  &--  &   & --  \\         
J033236.72-274406.4 &  03:32:36.715 $-$27:44:6.435 &  0.66500 & $-$20.23 & $-$22.01  &  58.7  & (1.8) &  UC &  &--  &   & --  \\   
\hline \hline														     
\end{tabular} 
\end{center}
\begin{list}{}{}
\item[$^{\rm{a}}$] Redshift measured by \oii\ emission.
\item[$^{\rm{b}}$] Absolute magnitudes in $B$- and $J$- band.
\item[$^{\rm{c}}$] Inclination of galaxies and its error (in unit of degree).
\item[$^{\rm{d}}$] Kinematical classification (see Sect.~\ref{seccls} for details): 
RD-rotating disks; PR-perturbed rotations; CK-complex kinematics; UC-unclassified.
\item[$^{\rm{e}}$] The corresponding error of each quantity is given
  into brackets. 
\end{list}
\label{tbobjects}
\end{table*}
%\end{sidewaystable*} 
}

\section{Kinematics of distant galaxies}
\subsection{Measuring galaxy kinematics using the \oii\ doublet }
\label{secMethod}
Our method to extract kinematic fields from 3D spectroscopy of the
{\sc [Oii]}$\lambda$3737,3729 emission line doublet has been described
in detail in Flores et al.\ (2006). Here, we only give a brief
overview, and highlight recent improvements. The 20 individual spectra
of each object were inspected visually to detect possible artefacts or
contamination with night sky lines. We then constructed the
3D data cube around the expected observed \oii\ wavelength with 
and without sky subtraction, and fitted the \oii\ doublet with two
Gaussian, keeping the wavelength difference between the two lines at
rest-frame $\lambda _2\!-\!\lambda _1\!=\! 2.783$\,\AA\ fixed, and
requiring that both lines have the same dispersion,
$\sigma_1\!=\!\sigma _2$. The line ratio is a free parameter except
when the fit failed, in which case we impose a ratio of R(3729/3727)
=1.4, which corresponds to the low density limit and is appropriate for most galaxies (see
also Puech et al.\ 2006b; Weiner et al.\ 2006).

We estimate the systemic velocity of each galaxy from the
$\sigma$-clipped mean of the spatially-resolved velocities, and
measure the width of night sky lines to correct the dispersion maps
($\sigma$-maps) for instrumental resolution. We also derive S/N-maps
to quantify the uncertainty of the kinematics following the definition
of Flores et al.\ (2006); in particular, we use only those spectra
where the \oii\ line emission is detected with a S/N $\ge3$, and apply a
simple 5$\times$5 linear interpolation to the VF and $\sigma$-map. We
show the VF, $\sigma$-maps and S/N-maps in Fig.~\ref{figkinematics}
with the high resolution ($0.03''$/pixel) ACS F775W image for each
object.  The analysis of the full sample was done independently by
several of us (HF, BN, MP and YY), before comparing and finalizing the
results.

Since Flores et al.\ (2006), we have improved our analysis software to
better account for the contamination of the emission line spectrum
with night sky lines, fitting the \oii\ emission lines and
superimposed night sky lines simultaneously. This is particularly
relevant with the L05 set-up (5741\,--\,6524\,\AA), where the risk of
overlaps is important due to a relatively large number of night sky
lines. At the relatively high effective resolving power of GIRAFFE of
R\,$\ge$\,10\,000, the \oii\ emission lines are significantly broader
than the night sky lines, which is essential to successfully isolate
the signal (see Fig.~\ref{figdecom} for an example). By fitting the
sky and object simultaneously, we have been able to recover the
kinematics of 3 galaxies that were particularly strongly blended with
night sky lines, and would have been otherwise rejected 
(J033217.62-274257.4, J033220.48-275143.9, J033244.20-274733.5).

\begin{figure*}
%\sidecaption
\centering
\includegraphics[width=14cm]{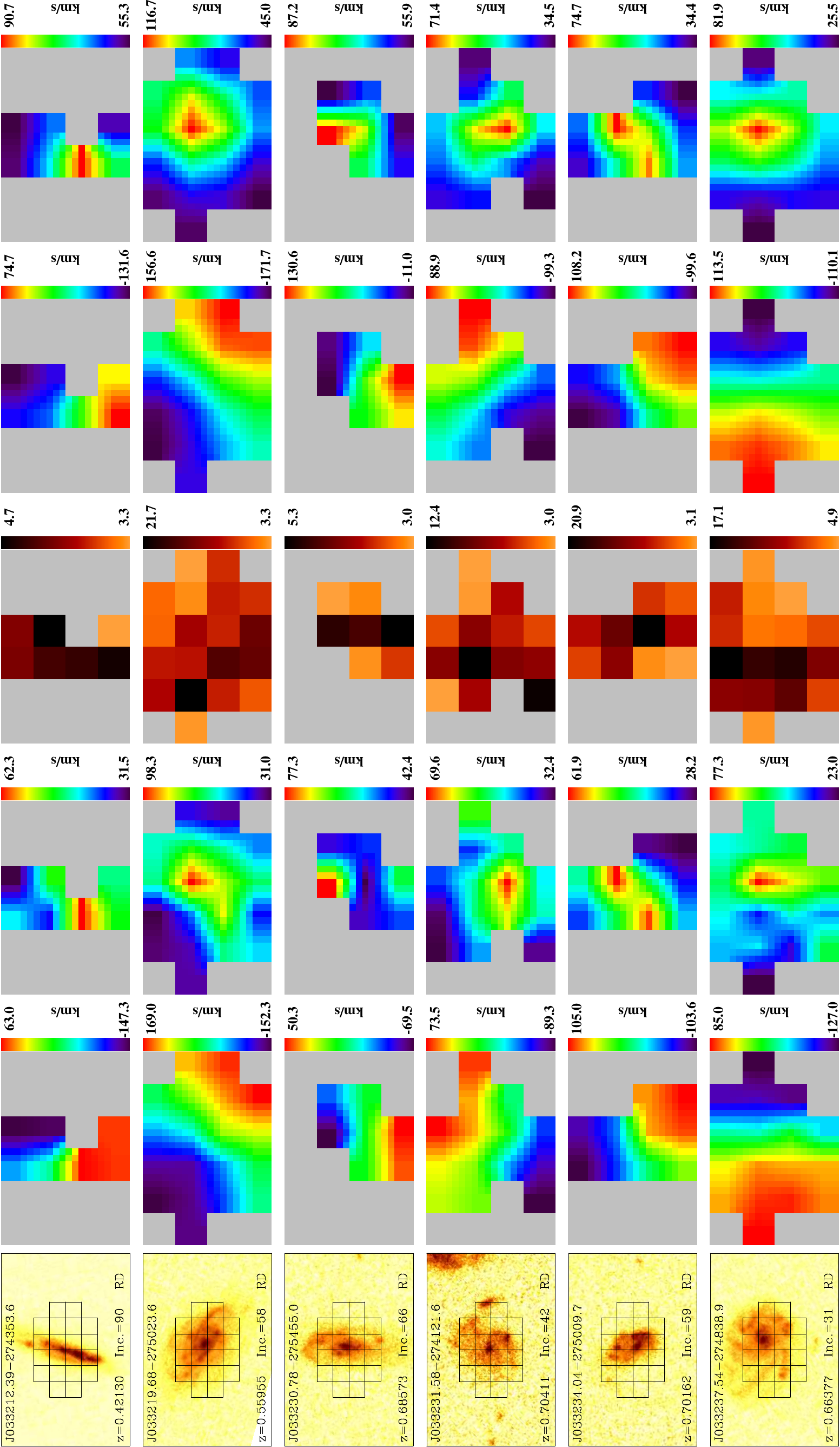}
\caption{Kinematics of the individual galaxies. Each row corresponds
to one galaxy. From left to right, we show the HST/ACS F775W images,
the observed VFs, $\sigma$-maps and S/N-maps, the model VFs and $\sigma$-maps. A grid
of GIRAFFE IFU superposed on the HST image indicates the position and
the scale of IFU with respect to the galaxy. We have applied a
5$\times$5 linear interpolation to the VFs and $\sigma$-maps for
visualization. }
\label{figkinematics}
\end{figure*}
\setcounter{figure}{1}
\begin{figure*}
\centering
\includegraphics[width=14cm]{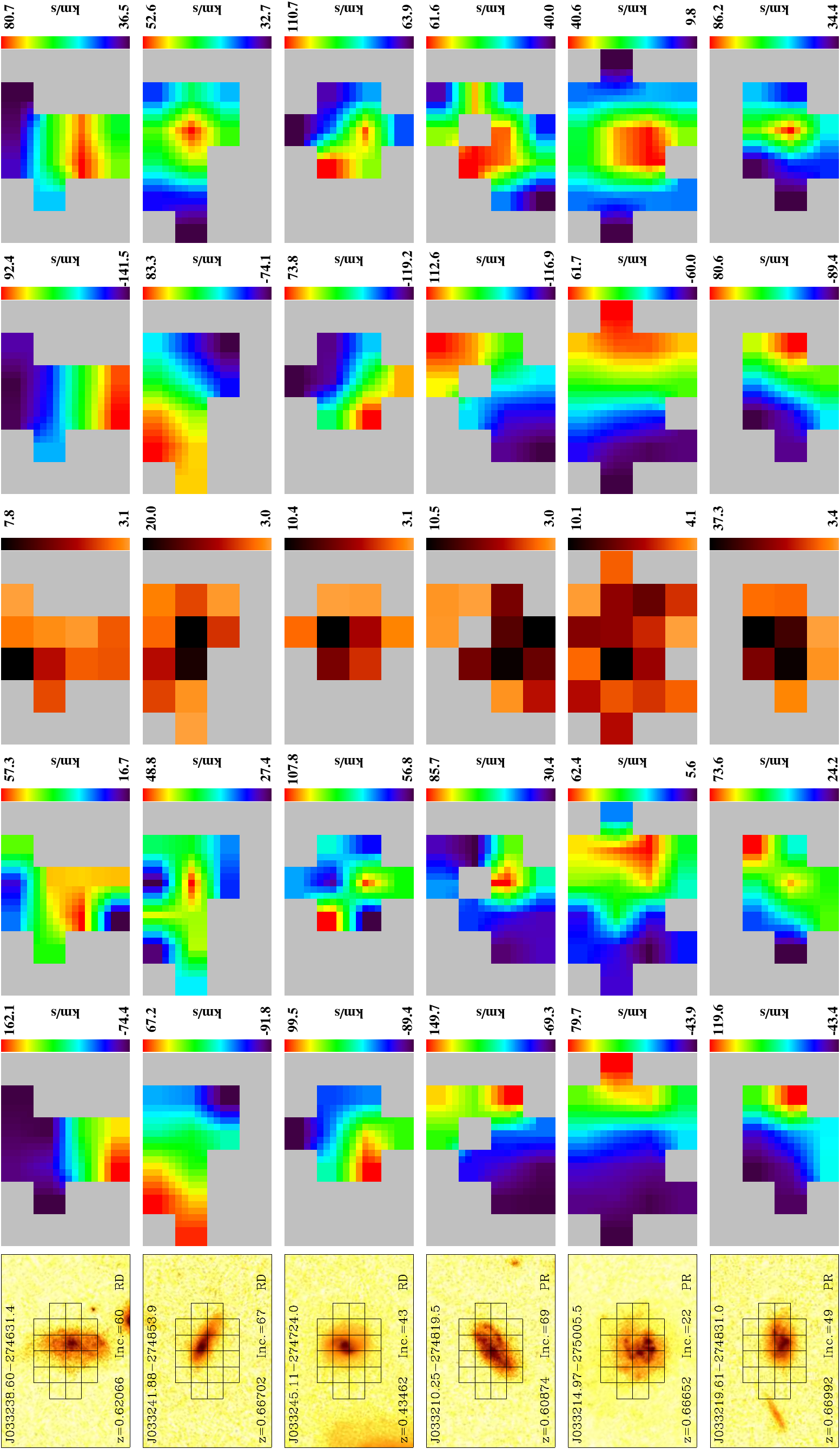}
\caption{continued.}
\end{figure*}
\setcounter{figure}{1}
\begin{figure*}
\centering
\includegraphics[width=14cm]{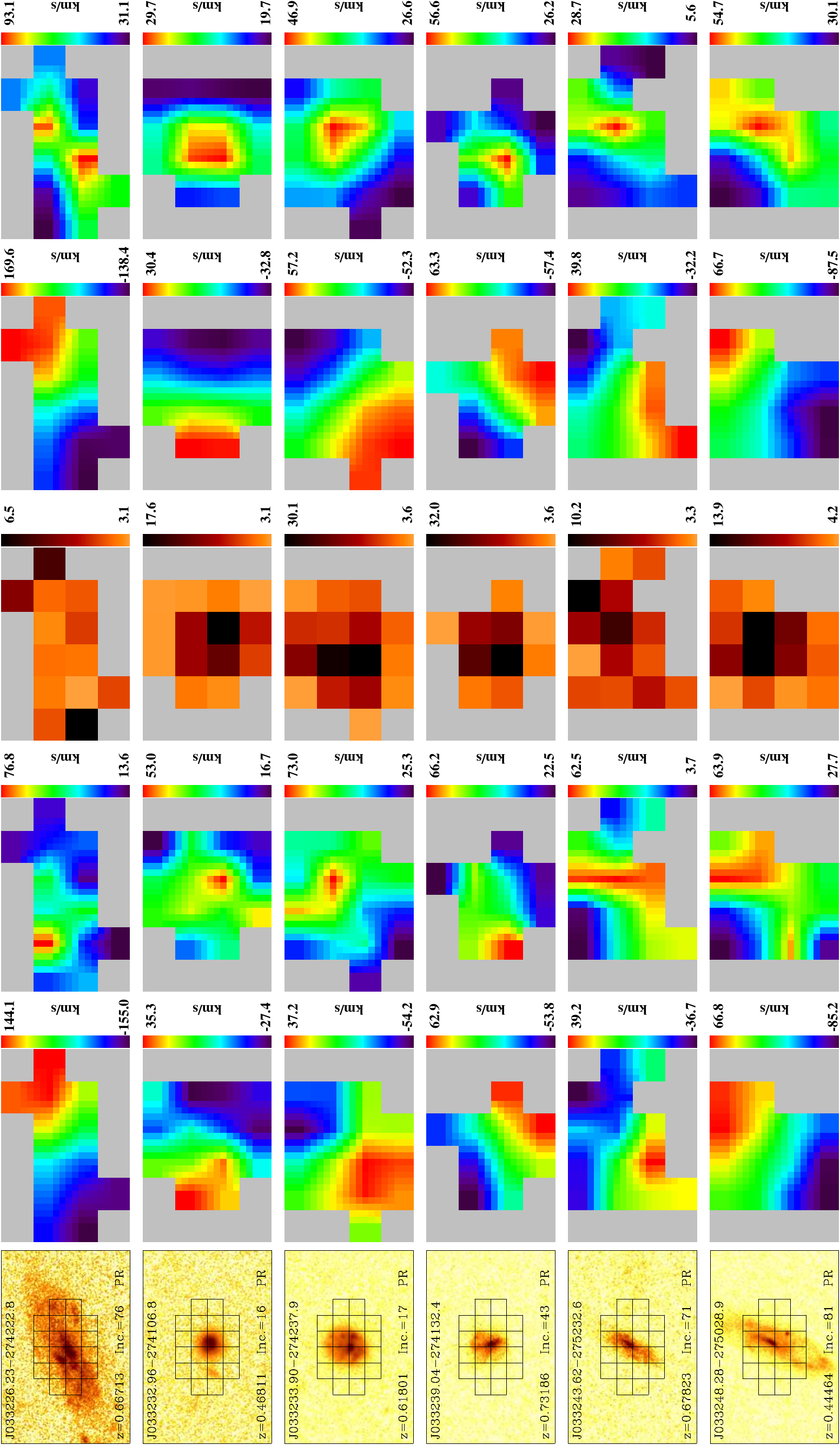}
\caption{continued.}
\end{figure*}
\setcounter{figure}{1}
\begin{figure*}
\centering
\includegraphics[width=14cm]{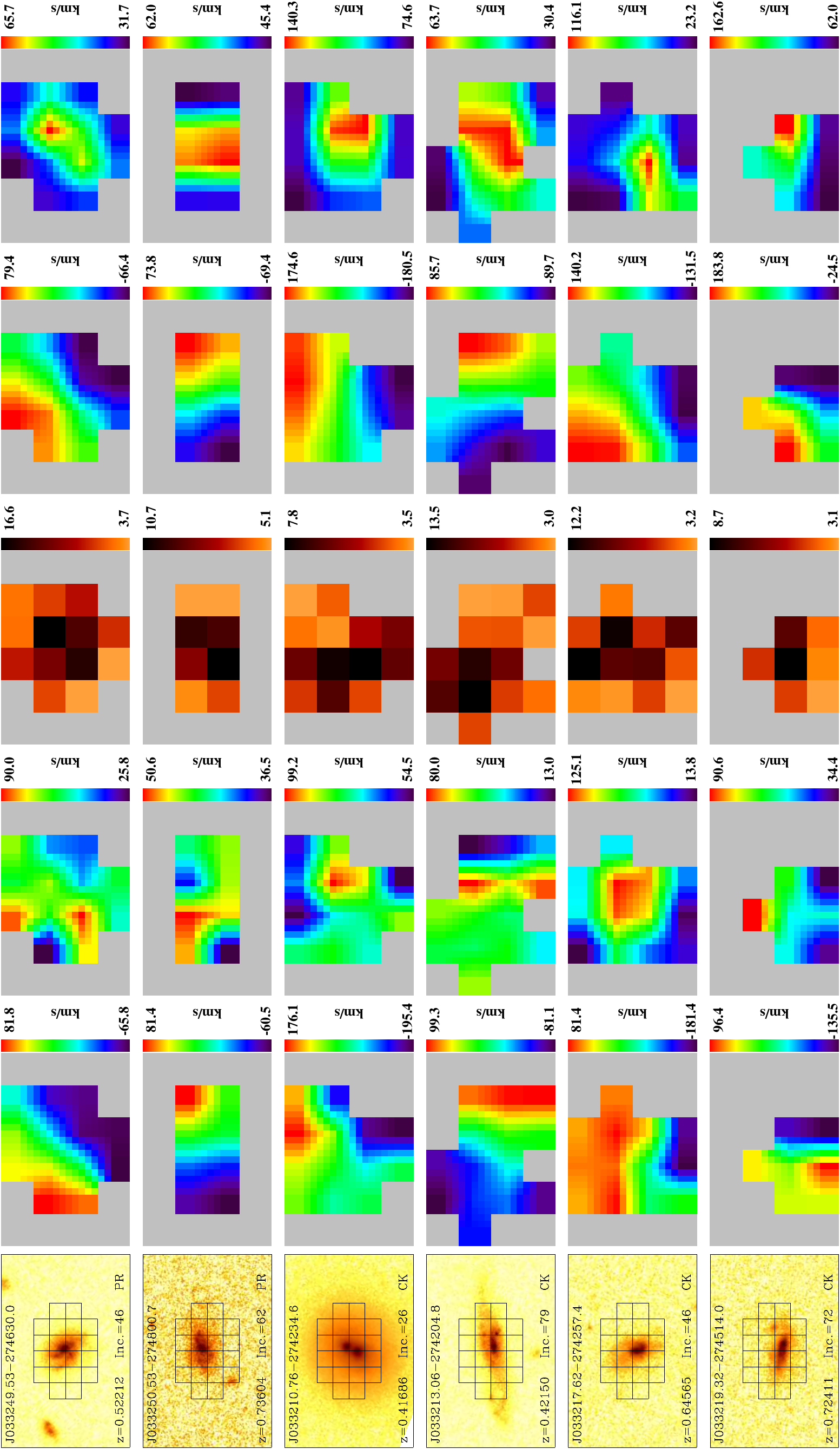}
\caption{continued.}
\end{figure*}
\setcounter{figure}{1}
\begin{figure*}
\centering
\includegraphics[width=14cm]{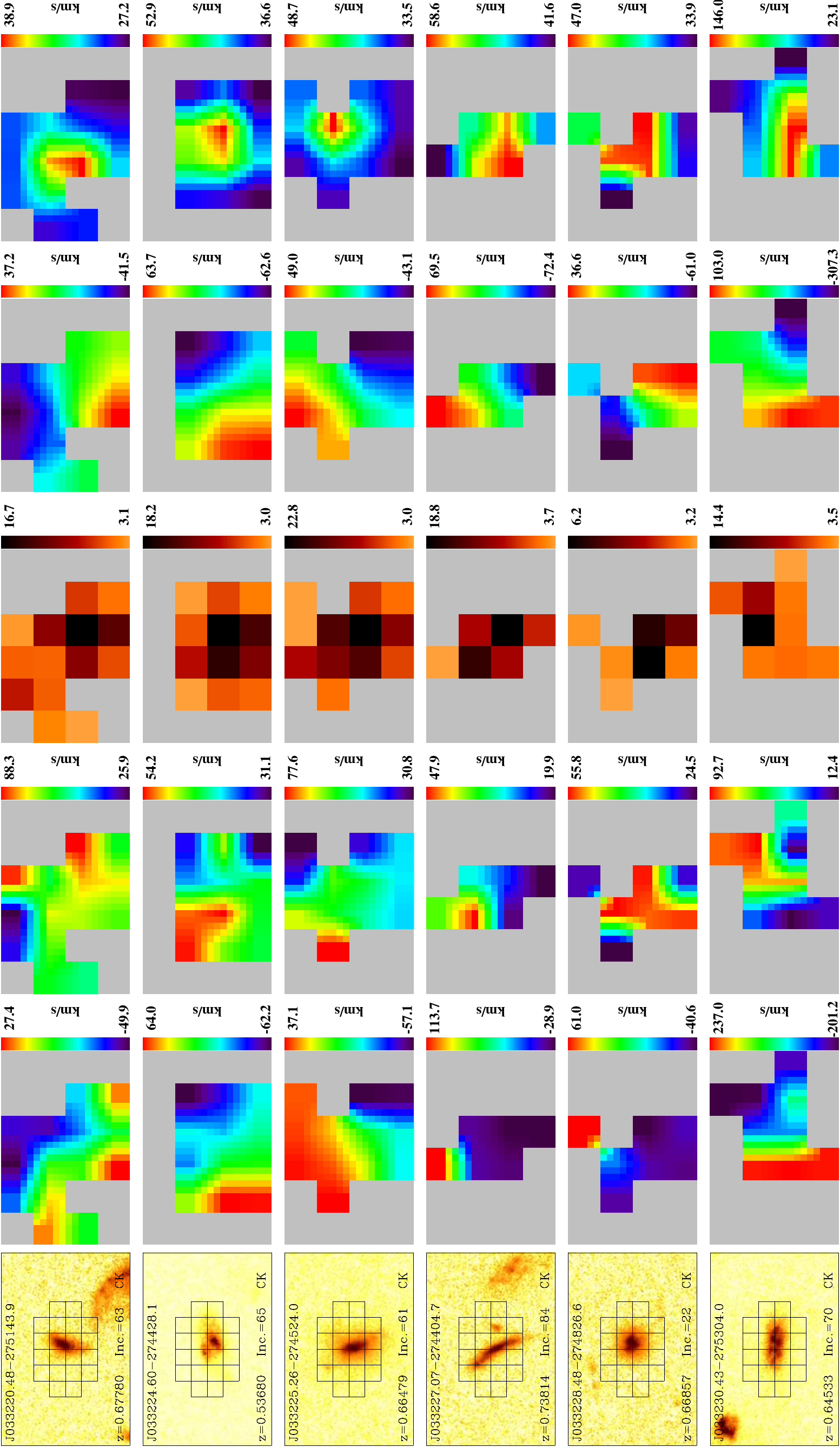}
\caption{continued.}
\end{figure*}
\setcounter{figure}{1}
\begin{figure*}
\centering
\includegraphics[width=14cm]{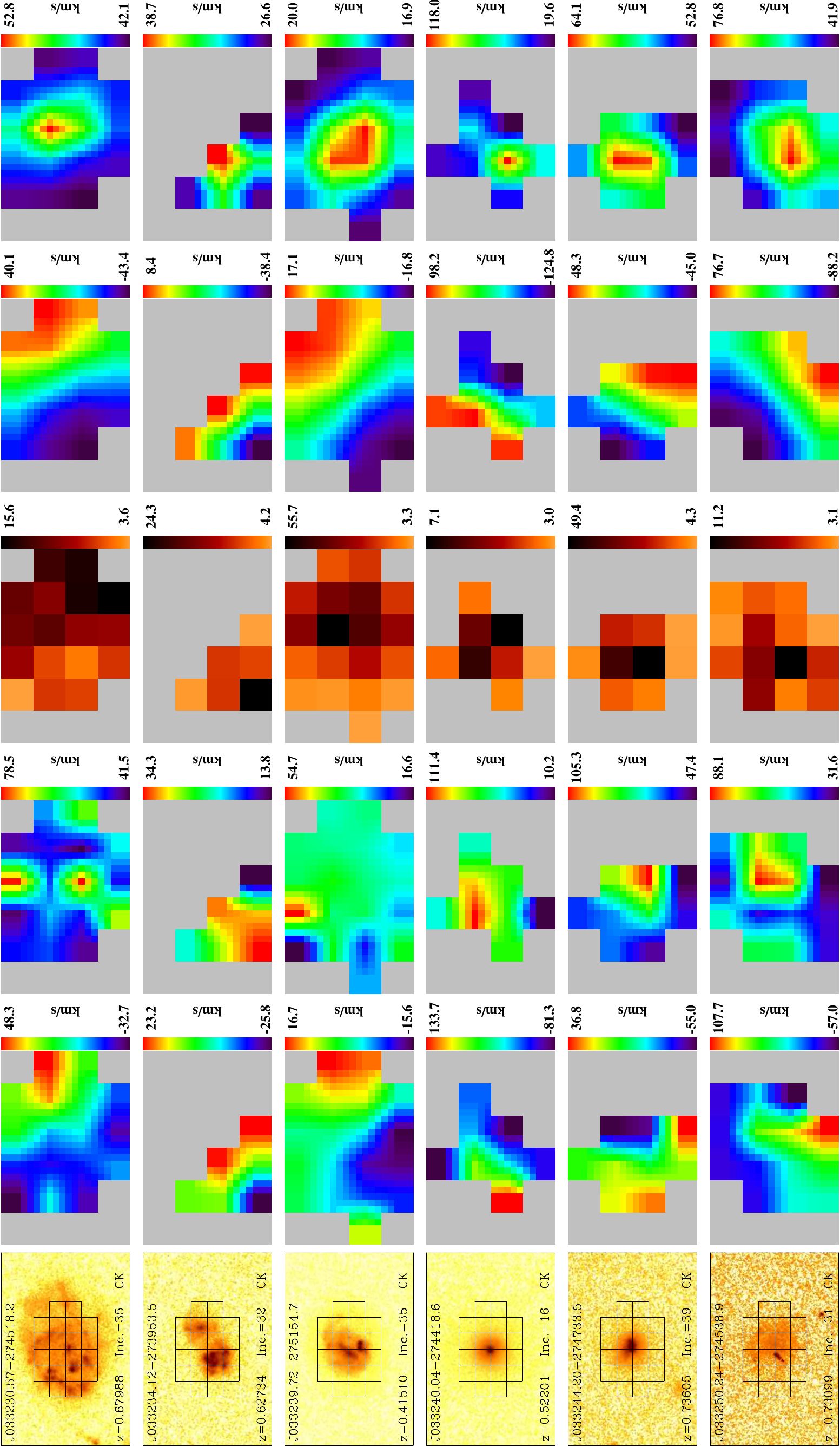}
\caption{continued.}
\end{figure*}

\begin{figure}[!ht]
\centering
\includegraphics[width=6.5cm]{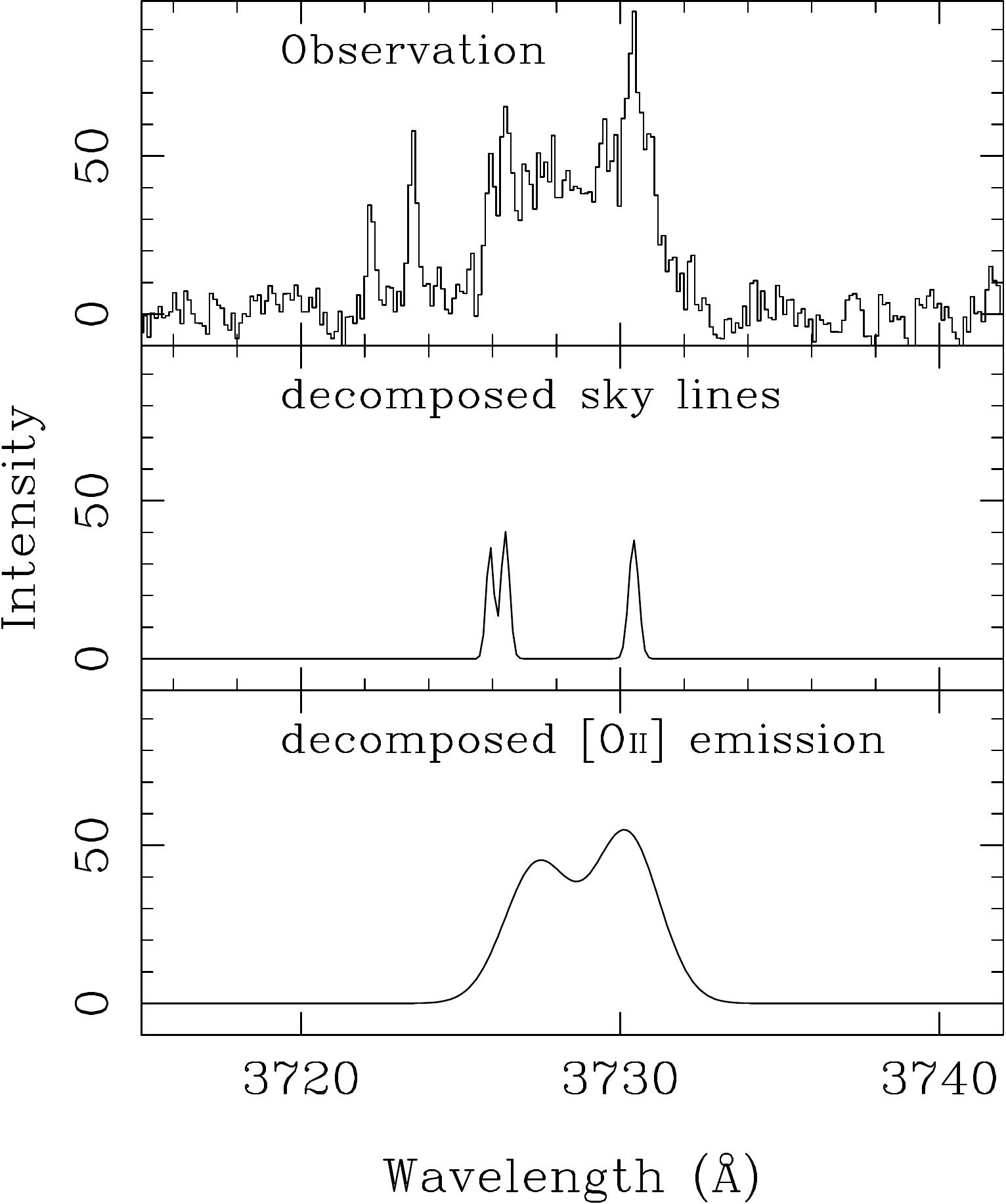}
\caption{Example of a simultaneous fit to the emission line of a
galaxy with superimposed night sky lines. The top panel shows the
observed \oii\ emission, which has three night sky lines
superimposed. The mid panel shows the isolated components of the night
sky lines after the fit, and the bottom panel shows the reconstructed
\oii\ emission line doublet.}
\label{figdecom}
\end{figure}

\subsection{ Classification of the kinematics of distant galaxies}
\label{seccls}
Flores et al.\ (2006) developed a simple kinematic classification
scheme for distant galaxies based on their 3D kinematics
and their morphology in the ACS F775W images. It relies on the fact
that at low spatial resolution, a rotating disk should show a well
defined peak in the center of the $\sigma$-map, which corresponds to the
convolution of the large scale motion (i.e., the rotation) with the
(relatively small) dispersion of the gas in the disk or in the
bulge. Indeed, the central parts of the galaxy, where the rotation
curve rises most quickly, are not spatially resolved with ground-based
optical spectroscopy and our classification fully accounts for its
convolution with the actual PSF (point spread function).  To
summarize, we distinguish between the following classes :
\begin{enumerate}
\item {\bf Rotating disks (RD):} 
the VF shows an ordered gradient, and the dynamical major axis is
aligned with the morphological major axis. The $\sigma$-map indicates a single
peak close to the dynamical center; 
\item {\bf Perturbed rotations (PR):} the kinematics shows all the
features of a rotating disk (see above), but the peak in the $\sigma$-map is either absent
or clearly shifted away from the dynamical center;
\item {\bf Complex kinematics (CK):} neither the VF nor
the $\sigma$-map are compatible with regular disk rotation, including
VFs that are misaligned with the morphological major axis.
\end{enumerate}

According to these definitions of 3 kinematical classes, we have
classified the 36 galaxies of our sample. To do so, each galaxy has
been examined by 5 of us (HF, FH, BN, MP and YY), before comparing our
classification. The results are listed in
Table~\ref{tbobjects}, which also includes absolute magnitudes and
inclinations. Absolute magnitudes are derived using the
procedure described by Hammer et al. (2005), based on photometry at near
IR and optical wavelengths (see also Ravikumar et al.\ 2007).
Inclinations were measured using ellipsoidal isophotes near the
optical radius from HST/ACS F775W and F814W imaging. Comparison
between estimates from different team members and with values
derived from Sextractor (Bertin et al.\ 1996) suggests that typical
uncertainties are about 5$^{\circ}$.

In some cases, classification is not an easy task. There are two
galaxies (J033234.04-275009.7, J033245.11-274724.0) that possess double
$\sigma$-peaks in GIRAFFE IFU view. They are classified to be RD
because the center of the galaxy is just located in between the two adjacent 
GIRAFFE IFU pixels. These phenomena have been reproduced in our model
(see Fig.~\ref{figkinematics}). However, another galaxy
J033219.61-274831.0, which is classified as PR, has its two
$\sigma$-peaks located at one side of the galaxy center, which is not
expected from a rotating disk.

In the next paragraphs, we will describe quantitatively the
differences between kinematical classes, a test of the robustness of
our classification scheme and a description of each individual target.

\subsection{A powerful diagnostic of the classification }
\label{secedplot}
\subsubsection{Measuring the discrepancy from a pure rotational disk }
\label{secmethod}
Flores et al.\ (2006) developed a diagnostic method to test the
validity of the classification, and to quantify the deviation of a
given VF from that of a pure rotational disk. First, we
identify the VF, based on the largest scale motion in each galaxy. We
then assume that this VF is the result of pure rotational motion,
whatever the true nature of the dynamics. We then model the VF of a
rotating disk that matches the observed velocity gradient estimated
from the measured minimal and maximal velocities to obtain the
expected $\sigma$-map corresponding to the observed VF (a
``model VF''). In particular, we take into account that most of
the velocity gradient in the central region (from 45\% to 70\%) falls
within one spatial pixel of GIRAFFE, as we observe in well-identified
rotating disks. To generate the model VF and $\sigma$-map shown
in Fig.~\ref{figkinematics}, we then use a single rotation curve that
concentrates an equivalent fraction of the velocity gradient in one
GIRAFFE pixel. While this should not affect the location of the
$\sigma$ peak, we are aware that this simplistic assumption may affect
the $\sigma$ peak intensity for some galaxies with flatter rotation
curves or with a prominent bulge. Finally, for model $\sigma$-map,
we assume that the intrinsic dispersion (due to the random gas motion
in the disk) is the smallest dispersion observed in the data.

The model VF for each galaxy has then been generated by assuming a
rotation curve that is been used to generate a model data cube. Note
that, whenever possible, we tried to align the model rotational axis
with the major axis of the galaxy, in agreement with the rotating disk
hypothesis.  Unlike what was done in Flores et al.\ (2006), we do not
try to correct for inclination, as the observed and the model VFs are
affected in a similar way by inclination effects. However, note that
the inclination is used during the process to define the geometrical
extent of the VF in the plane of the sky, assuming a thin
disk.  Further, we compute the corresponding IFU \sigmamap by
considering the effects of seeing. By comparing the observed and model
$\sigma$-maps, we can estimate whether the observed kinematics are
consistent or not with a rotating disk.

Two parameters are then computed to characterize the differences
between the two $\sigma$-maps, taking the model $\sigma$-map as
a reference. The first parameter is the spatial separation ($\Delta
r$, in GIRAFFE pixels) between the peaks in the two
$\sigma$-maps. This indicates how the center of rotation is recovered
by our observation. For each $\sigma$-map the pixel including the
$\sigma$ peak is identified, and then the peak location is calculated
as the barycenter of the surrounding pixels.  We verify that weighting
the barycenter by S/N does not change the result, so we choose a
uniform weight for each pixel. The second parameter is the relative
difference ($\epsilon$) between the amplitudes of the modeled and
observed $\sigma$ peaks. We define $\epsilon$ as:
\begin{equation}
\epsilon=\frac{|\sigma_{\rm obs}-\sigma_{\rm mod}|}{\sigma_{\rm mod}}
\bigg{|}_{\mbox{\scriptsize location of the peak of model \sigmamap}}.
\label{eqepsilon}
\end{equation}
This definition significantly improves the robustness of the test
compared with what was done in Flores et al.\ (2006), because the test
is now applicable to the pixels of highest S/N in
the center of the \sigmamap. One may wonder whether the test is
reliable, given the possible uncertainties due to our observational
set-up with low spatial resolution and limited S/N,
especially in the outskirts of the galaxies. In fact, some parts of
the VF of distant galaxies might have been missed, for example, when
the galaxy may extend further than the IFU. Alternatively, low
S/N near our cut-off limit may generate an absence
of detection in some extended parts of the galaxy. An illustration of
this is given by J033230.78-275455.0 (see Fig.~\ref{figkinematics}),
for which the S/N ratio is below the cut-off for almost half the disk
. This means that for some objects we may have missed part of the
rotation curve, or we may have introduced an artificial asymmetry. The
immediate consequence of this would be to generate a $\sigma$ peak
that is smaller in amplitude than the real one and/or that is slightly
offset from the rotational center.  However, the impact of the above
has to be quite marginal and has no effect on our classification after
a careful examination of individual VFs. Moreover, with our method to
build model rotational VFs, the maximal model rotation is produced by
the observed large scale motions, and we do use the same hot pixels
for both model and observed maps.

\begin{figure}[!ht]
\centering
\includegraphics[width=7cm]{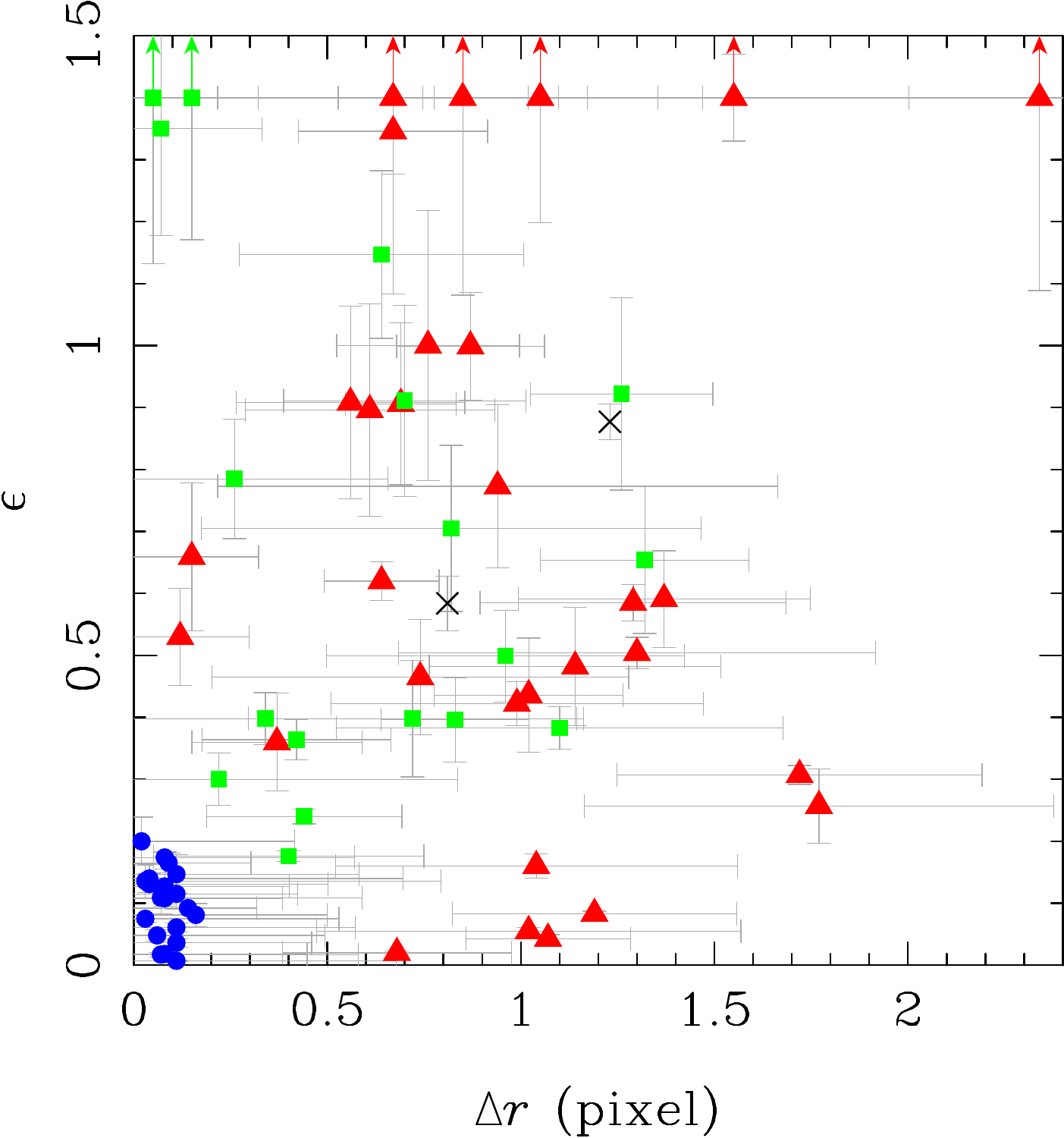}
\caption{Diagnostic diagram of the galaxy kinematics. $\Delta r$ is
the spatial separation between the peaks of the modeled and the
observed $\sigma$-maps. $\epsilon$ gives the relative difference
between the velocity dispersions of the model and the observation at
the reference location of the peak of the model $\sigma$-map. The blue
dots, green squares and red triangles represent the RD, PR and CK
galaxies, respectively. Typical uncertainties are 0.3 spatial pixels
in $\Delta r$, and 15\% in $\epsilon$. (see Sect.~\ref{secedplot} for
more details.)  }
\label{figedmod}
\end{figure}

Figure \ref{figedmod} shows the diagnostic diagram of $\Delta r$
versus $\epsilon$.  We have recalculated $\epsilon$ for the 32
classified galaxies from Flores et al.\ 2006, and plotted them in the
same figure. While modifying the definition clearly changes the
location of the points, we confirm all the results of Flores et al.,
i.e., that rotating disks have locations near $\Delta
r\!\sim\!\epsilon\!\sim\!0$, while galaxies with anomalous VFs are
clearly offset. \new{We have performed simulations of two local
interacting pairs: ARP\,271 and KPG\,468 (Fuentes-Carrera et al.\
2004; Hern{\'a}ndez-Toledo et al.\ 2003), which are supposed to be CK
systems. We assume that they are located at z=0.6 and observed by GIRAFFE
IFU mode in the simulations. Following our method, we find these two
simulated CKs (indicated by black crosses) are located
far from the rotating disks, supporting our diagnostic method.}

\subsubsection{Using $\chi^2$-test to recognize disk kinematics}
\label{secchi2}
\begin{figure}[!ht]
\centering
\includegraphics[width=7cm]{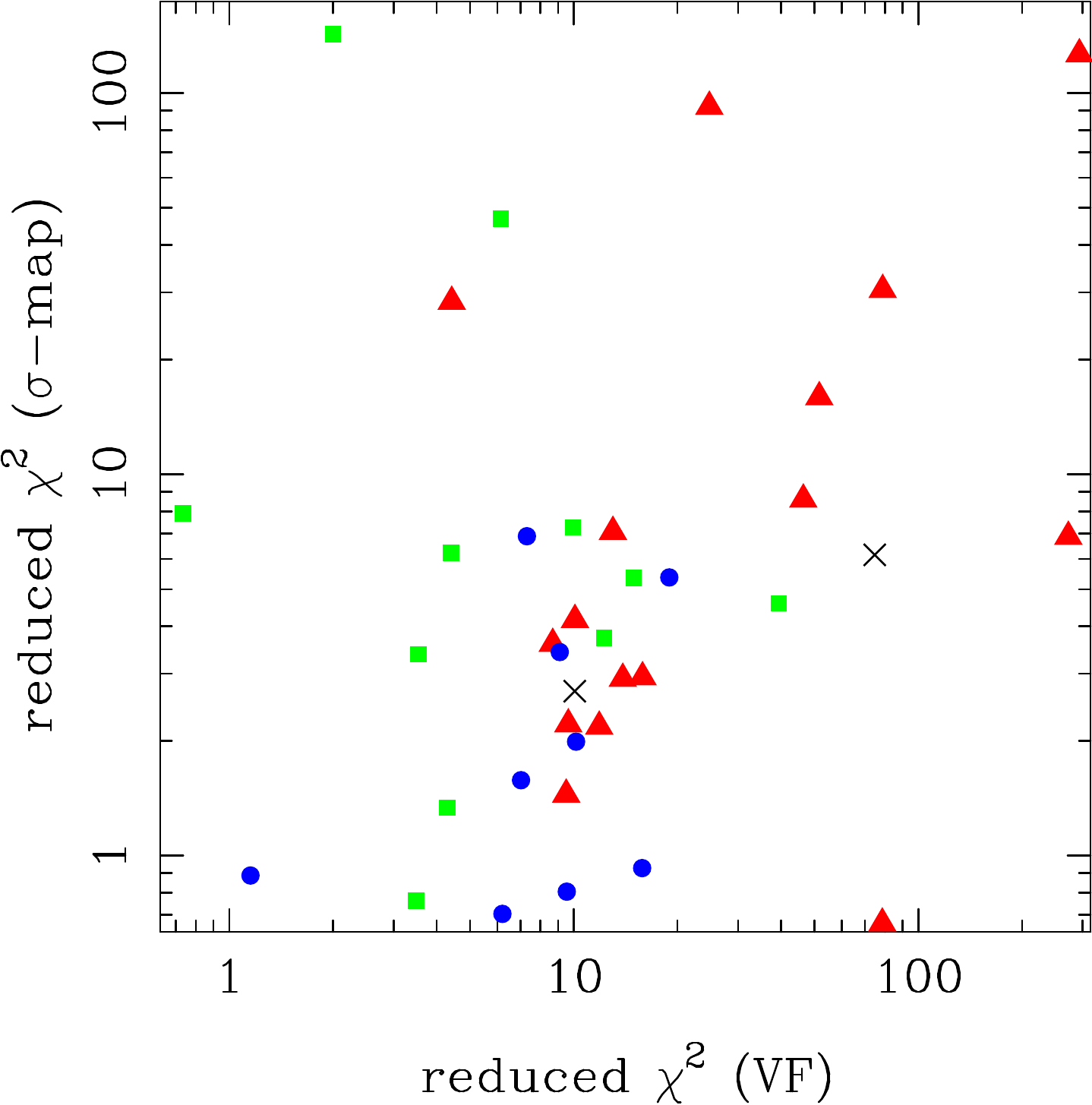}
\caption{The $\chi^2$-diagram for the best-fitting rotation model
of galaxy kinematics. The reduced $\chi^2$ for VF-map and $\sigma$-map
are set to be abscissa and ordinates, respectively.  The blue dots,
green squares and red triangles represent the RD, PR and CK galaxies,
respectively. The two black crosses indicate the two simulated galaxy
pairs: ARP271 and KPG468 (see Sect.~\ref{secmethod} for
more details).  Although the CKs tend to have higher
$\chi^2$ than the PRs and the RDs, it is
still less efficient to identify different kinematics with this
diagram. (see Sect.~\ref{secchi2} for more discussion.)}
\label{figchi2}
\end{figure}

Given the spatial resolution of IFU and the seeing during
observations, we only have a few degrees of freedom to confine the
model. Hence, $\chi^2$-test may not be an ideal tool for comparing the
model and the observations. A series of tests have been performed in order
to explore the validity of the $\chi^2$ estimator of goodness of the
fits. In Fig.~\ref{figchi2} we present a best-fitting $\chi^2$-diagram
for the CDFS sample.  First, we fit the observational VF-map with a rotation
model, then we calculate the reduced $\chi^2$ which is set as
abscissa.  The corresponding $\sigma$-map is generated following the
same method described in the previous section, then the reduced
$\chi^2$ for the $\sigma$-map is computed and presented as ordinates.
%In this figure, blue dots, green squares and red triangles represent
%the RD, PR and CK galaxies, respectively.  
The two simulated galaxy
pairs, ARP271 and KPG468 (see the caption of Fig.~\ref{figedmod} for
more details), are also plotted.  Although, the CK
systems tend to have larger $\chi^2$ than the RD and PR, it is
difficult to distinguish different kinematical classes with such
diagram.

Taking into account the spatial resolution of the GIRAFFE IFU, the
most important features of a rotating disk are (a) a well ordered
VF with dynamical axis following the optical major-axis of
the galaxy, (b) a clear $\sigma$-peak located at the position of the
galaxy center, (c) and the $\sigma$-peak having a reasonable value, which
can be generated by rotation. These important features are somehow
diluted by the $\chi^2$-test as it is presented in
Fig.~\ref{figchi2}. For example, J033230.78-275455.0, a RD
galaxy, has a reduced $\chi^2$ close to 1 in both VF- and $\sigma$-
maps because of its relatively low averaged S/N ($=4$); other galaxies
(which have typically averaged S/N $\sim 8$) have $\chi^2$ of $\sim$
10. The $\chi^2$-test appears to be too sensitive to the averaged
S/N. Furthermore, using this type of test we cannot distinguish whether one of
the simulated galaxy pairs (KPG468) is a  RD or a PR, since the
$\chi^2$ value of this pair is similar to that of both RDs and PRs.
Since the $\chi^2$ values are to some extent scaled by the errors
that are related to the S/N, we also tried to apply an unweighted
$\chi^2$-test. In this case, we still have the same disordered
distribution over the $\chi^2$ diagram.

Consequently, the $\chi^2$-test seems less
efficient than our method (see Sect.~\ref{secmethod} and
Fig.~\ref{figedmod}) in recognizing disk kinematics
from perturbed and complex kinematics.  Therefore, 
we choose to compare the observations with the rotating disk model using
the methodology developed in Sect.~\ref{secmethod}.

\subsubsection{Error estimates}
We used Monte-Carlo simulations to evaluate the uncertainty of
our velocity dispersion measurements. First, we generate $\sim$50\,000 %a set of
artificial \oii\ emission line doublets with S/N
ranging from 3 to 40. The flux ratio within the doublets is fixed at 1.4.  
Simulated dispersions range from 10 to 100 km s$^{-1}$, corresponding
to the minimum and maximum of our observations, respectively.

In order to remove possible artefacts in the data, we analyzed the
spectrum of each spatial pixel manually in each data cube before
thoroughly analyzing the spectral fit of each individual line. To
ensure that the Monte-Carlo simulations are a fair representation of
this process, we did not only consider the results from automatic
fitting routines, but use the same procedure that was used for the observed
data-cubes. This is particularly important for spectra with low
S/N (i.e., 3\,--\,5) spectra, and it somewhat
reduces the uncertainties of our measurement.

\begin{figure}[!ht]
\centering
\includegraphics[width=7cm]{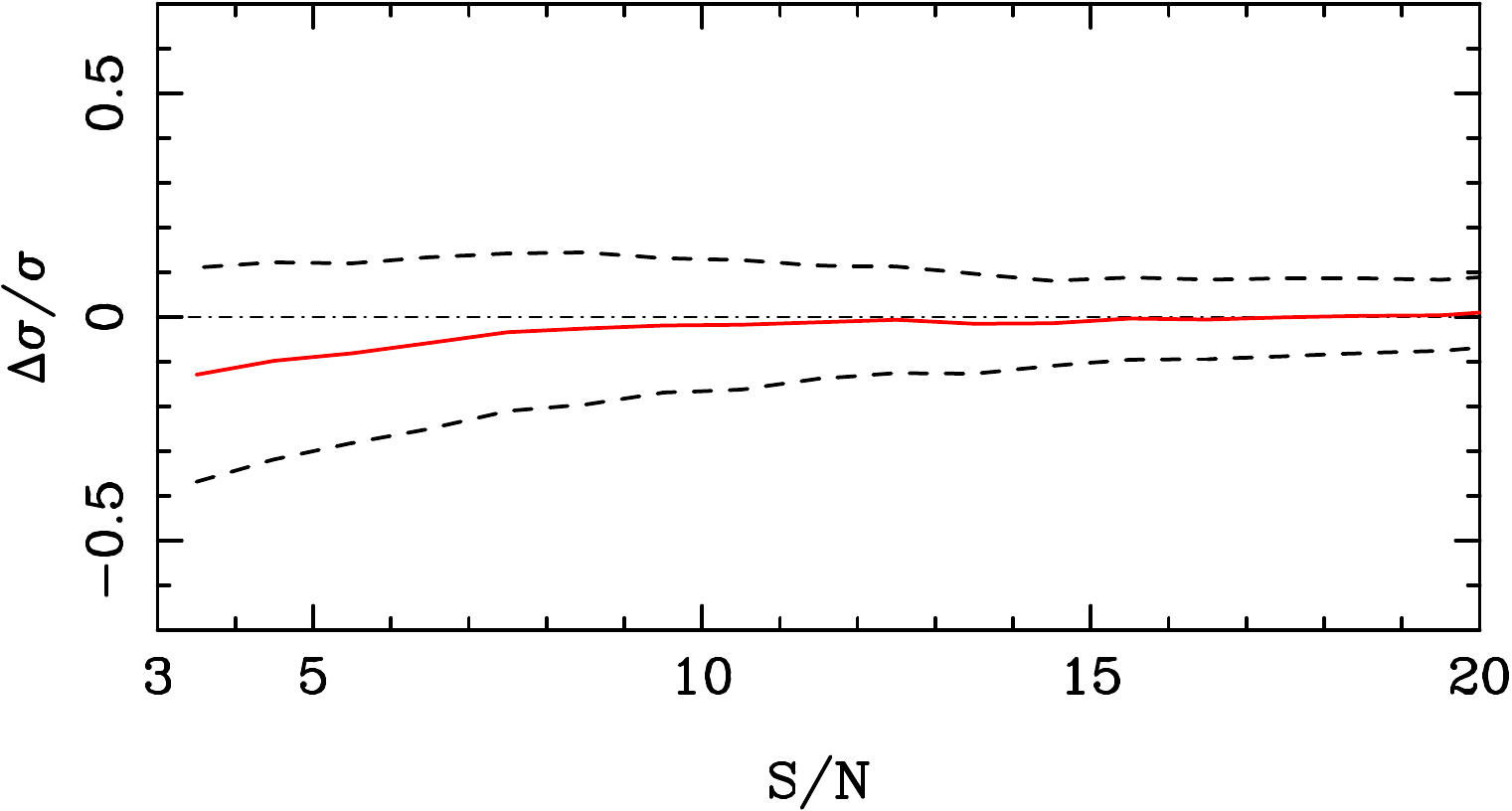}
\caption{The red solid line indicates the median
differences between the measured and the simulated $\sigma$ values
varying with the S/N. The two dashed lines give the
1$\sigma$ errors for each bin in S/N: typically,
these amount to $\sim$20\% at $3\!<\!{\rm S/N\!<\!4}$, 15\% at ${\rm
S/N\!\sim\!10}$ and about 5\% at ${\rm S/N\!>\!15}$, respectively. }
\label{figerr}
\end{figure}

We then investigated the uncertainties in the velocity dispersion as a
function of the S/N. Figure \ref{figerr} shows the relative
uncertainty in the velocity dispersion ($\Delta \sigma/\sigma$) as a
function of the S/N. \new{We find a slight trend that we possibly
underestimate velocity dispersion when we have spectra with S/N lower
than $\sim5$, because the noise affects the line wings in
particular. However, this systematic uncertainty is negligible
compared to the statistical error.} Based on this result, we estimate
the uncertainty of the observed velocity dispersion at a given S/N.
Standard error propagation then yields the uncertainty on $\epsilon$
(Eq.~\ref{eqepsilon}). Typically, we find that uncertainties are $\sim
15$\%, ranging from 5\% to 25\% (see Table~\ref{tbobjects} for the
uncertainties of individual objects).

Uncertainties in $\Delta r$ are dominated by the error on the location
of the peak of the velocity dispersions. This uncertainty is dominated
by the profile of the \sigmamap and its S/N. Again, we use Monte-Carlo
simulations to quantify these uncertainties. We calculate the
uncertainty of each value in our $\sigma$-maps using the S/N-map and
Fig.~\ref{figerr}. As a next step we generate a set of simulated
$\sigma$-maps from the observed maps by randomly shuffling the
$\sigma$ values within the error distribution, and derive the
uncertainty in the peak location from the statistics of the
Monte-Carlo simulations.

We thus find that the largest uncertainties are related to the spatial
sampling of GIRAFFE, i.e., the sub-pixel position adopted for $V_{\rm
max}$ and $V_{\rm min}$. In the models this yields a geometrical error
of about 0.14 pixels in the location of the peak.  For some very
extended galaxies (e.g., J033230.78-275455.0 and J033226.23-274222.8),
parts of the galaxy fall outside of the IFUs; in these cases, we
extrapolated the position of $V_{\rm max}$ and/or $V_{\rm min}$
according to the partial VFs and the optical images. Here,
we obviously underestimate the corresponding errors. However, most
galaxies in the sample are smaller than the field of view of an
individual GIRAFFE IFU and are fully sampled; since we are mainly
concerned about the ensemble properties of our sample, we did not
include this into the total error budget.  In total, the uncertainties
in the observed \sigmamap and those found from the models together
correspond to an error of 0.14 to 0.8 spatial pixels on $\Delta r$,
with a median of $\sim 0.3$ pixels (see Table~\ref{tbobjects}).

The clear distinction between rotating disks and more complex kinematics may
serve as additional evidence for the validity of our classification
scheme. Galaxies classified as rotating disks are well concentrated near the 
 $\Delta r\!\sim\!\epsilon\!\sim\!0$ region, which implies that they are well 
modeled by a simple rotating disk. Galaxies
classified as PR or CK fall outside of this region, and with large
scatter. Note that 
there are five objects with $\Delta r\sim0$ and $\epsilon>1$. Their
VFs resemble that of a rotation disk, but the amplitude of
their $\sigma$-peaks show a significant deficiency. Three of them,
(J033243.62-275232.6, J033219.32-274514.0 and J033234.12-273953.5) have an obviously disturbed $\sigma$-map and
VF, which is why we classify them as having perturbed or complex kinematics.
The two remaining galaxies (J033233.90-274237.9 and CFRS031032) are
dominated by bulges that are much bluer
than those of present day galaxies and presumably are experiencing
star formation (see Neichel et al.\ 2007). The presence of a star forming
bulge may affect the central velocity dispersion in such a way that a pure
rotating disk model underestimates the true $\sigma$-peak amplitude. It is then
possible that these two galaxies are indeed supported by the
combination of a rotation and dispersion, as expected for S0 galaxies.

During the final phase of the classification process (see
Table~\ref{tbobjects}), we have indeed used Fig.~\ref{figedmod} to
verify our results. It had led us to change one galaxy from PR
to RD class (J033230.78-275455.0, see discussion above) and one galaxy
from RD to PR class (J033248.28-275028.9), thus evidencing that
classification errors are marginal. Furthermore, one PR galaxy
(J033219.61-274831.0) falls close to the region of rotating disks and
has relatively large error bars in $\Delta r$ (see Fig.~\ref{figedmod}
and Table~\ref{tbobjects}).  Its $\sigma$-peak is located near the
edge of the $\sigma$-map, which causes a relatively large uncertainty
on the peak location. However, given the S/N of the
$\sigma$-peak and the presence of a secondary peak near the galaxy
center, there is no doubt about its classification as PR.

\subsection{Comments on individual targets}
\label{ssecindividuals}
\noindent {\bf J033212.39-274353.6:} Kinematically classified as a RD. 
The dynamical major axis is aligned with the morphological major
axis. The $\sigma$-map shows a clear peak near the center. \\ 
\noindent {\bf J033219.68-275023.6:} This is an excellent example of a
RD. The velocity gradient and the sigma peak are well consistent
with a RD, while the morphology clearly shows spiral arms.\\ 
\noindent {\bf J033230.78-275455.0:} It has the morphology of a spiral
galaxy.  Its \oii\ emission is relatively faint, and only the lower half of
the galaxy was detected.  Nevertheless, the peak of the \sigmamap is well
centered on the nucleus of the galaxy. The \sigmamap is in agreement with
our simple rotational model, where the center and PA have been chosen according to
the geometry of the optical image. Thus, we classify its VF to be a RD. \\ 
\noindent {\bf J033231.58-274121.6:} The kinematics indicates that this is a
RD, in spite of its irregular morphology (see Neichel et al. \ 2007). \\  
\noindent {\bf J033234.04-275009.7:} We find a well ordered
velocity gradient consistent with disk rotation, and dispersion peaks
near the morphological and kinematic center. Notice that the $\sigma$
peak falls just between two GIRAFFE pixels. This galaxy is classified
as a RD, in spite of its irregular morphology (Neichel et
al.\ 2007). \\
\noindent {\bf J033237.54-274838.9:} Kinematically, this galaxy appears to be
a RD. Morphologically, we find a nearly face-on disk galaxy with
an asymmetric outer region and prominent arms. Both the optical images and
kinematic maps show the typical features of a rotating system.  \\ 
\noindent {\bf J033238.60-274631.4:} The VF shows the characteristics of a
RD. Morphologically it has been classified  as an S0 galaxy. Thus,
we classify this galaxy as a RD.\\ 
\noindent {\bf J033241.88-274853.9:} A RD with a dynamical axis
 centered on the luminous peak of the red HST/ACS image. The morphology
 appears however asymmetric, and this galaxy has been classified as
 peculiar/tadpole by Neichel et al.\ (2007). Indeed, the ionized gas detected
 by the IFU has no stellar counterpart on one side of the galaxy.\\ 
\noindent {\bf J033245.11-274724.0:} We classify this galaxy as a RD,
in spite of a secondary peak, which is offset from the dispersion map.
However, this is a compact object with symmetric spiral arms.  The
double peak in the dispersion map can be reproduced by our simulation
and may be an artefact caused by the relatively low spatial resolution
of the data, due to the compactness of the source.\\
\noindent {\bf J033210.25-274819.5:} The VF is not well aligned with the
  major axis of the galaxy because of one high S/N pixel
  with the highest velocity. The $\sigma$ peak is found to be close to this
  pixel, and cannot be reproduced by rotation along the major
  axis, implying that the rotation is perturbed.\\ 
\noindent {\bf J033214.97-275005.5:} The peak
of the \sigmamap significantly deviates from the center of this face-on
galaxy, we therefore classify it as a PR.\\ 
\noindent {\bf J033219.61-274831.0:} Another PR for
which the \sigmamap shows a peak that is well offset from the
dynamical center. Morphologically, this appears to be a
peculiar, perhaps merging galaxy (Neichel et al.\ 2007). In the spectrum with
the largest velocity dispersion, we identify a narrow component superimposed
onto a broad component. This may be related to merging.
\\
\noindent {\bf J033226.23-274222.8:} The VF of this galaxy is classified as 
PR. The VF shows the
dynamical axis parallel to the morphological major axis. However, the peak in
the \sigmamap is not well aligned with this axis. This perturbed component in
the \sigmamap corresponds to a blue clump in the HST data (Neichel et al.\ 2007), which might be a gas rich satellite galaxy.\\ 
\noindent {\bf J033232.96-274106.8:} The peak in the \sigmamap is offset from 
the dynamical axis. Thus, this is clearly a PR. An unresolved
component near the bottom left corner of the kinematic maps may correspond to
a small companion of this galaxy, which perhaps is the cause of the
perturbation. \\ 
\noindent {\bf J033233.90-274237.9:} The \sigmamap shows a peak
near the dynamical center. Our rotating disk model can reproduce the
$\sigma$-peak in position but not its amplitude.
The morphological analysis by Neichel et al.\ (2007) suggests that this is a
S0  galaxy with a significant bulge (B/T=0.8). Our purely rotating disk
model does not take into account the bulge: if we add a bulge with a
typical velocity dispersion of $\sim$\,50\,km\,s$^{-1}$, then we are able to
reproduce the data. Conservatively, we classify this galaxy as a
PR, although the perturbation may simply be caused
by the prominent bulge.\\
\noindent {\bf J033239.04-274132.4:} The dynamical and morphological major
axes are aligned, but the peak of the \sigmamap is strongly offset from the
this axis.  Its VF is obviously a PR. The center of this
galaxy shows very luminous \oii\ line emission. \\ 
\noindent {\bf J033243.62-275232.6:} Classified as a PR
because the \sigmamap does not show a single peak but a very extended large
$\sigma$ region, which cannot be reproduced with our rotating disk model. \\ 
\noindent {\bf J033248.28-275028.9:} The VF is classified as PR. 
The \sigmamap shows an elongated peak across two spatial pixels,
and has a secondary peak. The extended peak of the \sigmamap cannot be reproduced
by our model of a rotating disk.  The overall blue colour (Neichel et al. \
2007) of the galaxy may be related to enhanced star-formation as a result of
the perturbed gravitational potential.\\ 
\noindent {\bf J033249.53-274630.0:} The VF of this galaxy is classified as 
PR because of the irregularity of its $\sigma$-map, which cannot
be reproduced assuming pure rotation. The morphological analysis by Neichel et
al.\ (2007) suggests that this may be a merger. \\ 
\noindent {\bf J033250.53-274800.7:} We detect that the dynamical axis
follows the major axis of the galaxy, but the peak of the \sigmamap
is clearly offset from the center. \\
\noindent {\bf J033210.76-274234.6:} It is classified to be CK without
evidence for a dynamical axis.  The peak of the \sigmamap is not
located at the center of the galaxy. \\
\noindent {\bf J033213.06-274204.8:} It has possibly a spiral
morphology (maybe with a bar), but the kinematics are complex. We find
the dynamical axis is clearly misaligned with the optical major axis.
The \sigmamap is irregular without any peak at the center,
and with an overall relatively high velocity dispersion
($\sim$\,50\,km\,s$^{-1}$) over the whole field.\\ 
\noindent {\bf J033217.62-274257.4:} The dynamical axis is misaligned
with the major axis, implying complex kinematics.  We also
detected that a very broad and high dispersion $\sigma$-peak
covers the majority of the galaxy.\\
\noindent {\bf J033219.32-274514.0:} Its kinematics is complex. Both VF and
\sigmamap are perturbed. No dynamical axis can be determined from the VF. \\ 
\noindent {\bf J033220.48-275143.9:} It clearly shows a CK,
from both VF and \sigmamap patterns. \\ 
\noindent {\bf J033224.60-274428.1:}  A complex VF, which shows a velocity
gradient along one component major axis while the \sigmamap shows a very large
region of high dispersion region that deviates from the center of galaxy. Its
morphology resembles that of an on-going merger.\\ 
\noindent {\bf J033225.26-274524.0:} The VF
shows a well ordered velocity gradient but skewed. Its \sigmamap does not show
any peak near the center. Thus, it was classified to be a CK.\\
\noindent {\bf J033227.07-274404.7:} We detect a very small
velocity gradient over most of the galaxy, 
with an amplitude of less than 20 km/s. The top end
clump in the optical image is responsible for the highest velocity
detected. The peak of the \sigmamap is offset from the center of
the galaxy and corresponds to the maximum gradient in the VF. Its VF and
coma-like morphology make us suspect that this is a merger between two or three
galaxies.\\ 
\noindent {\bf J033228.48-274826.6:} Both VF and \sigmamap show the signs of 
perturbation. No clear dynamical axis can be found. Its morphology is
classified as irregular. \\ 
\noindent {\bf J033230.43-275304.0:} Its VF is classified CK,
because the dynamical axis is not oriented along the major axis and the VF
does not show a clear velocity gradient consistent with rotation. The peak of
the \sigmamap is strongly offset from the center of the galaxy. \\  
\noindent {\bf J033230.57-274518.2:} It is classified as possessing complex
kinematics. Its VF shows a very asymmetric gradient while the \sigmamap shows
a clear difference from a single peak pattern. \\  
\noindent {\bf J033234.12-273953.5:} This galaxy has CK. 
We have detected a narrow VF and a low velocity dispersion for
this object. The peak of the \sigmamap is not corresponding to the center of
the galaxy.  It has an irregular morphology and a nearby companion, which
we do not detect in the \oii\ line emission. This may be an on-going merger or
a simple projection of two objects at different redshift. We also notice that
VVDS misidentified the \oii\ emission of this galaxy.\\ 
% does this mean the oii emission is near the end of the giraffe bandpass?
\noindent {\bf J033239.72-275154.7:} This is a galaxy with CK
showing no evidence for rotation. Its dynamical axis seems
to be parallel to a possible bar-like structure but it is clearly
offset from this structure. Because of its low inclination, it is
not expected to have a large amplitude of the velocity field. The \sigmamap
shows a peak clearly deviating from the galaxy center.  \\
\noindent {\bf J033240.04-274418.6:} The kinematics are classified as
complex. Its VF is perturbed and its \sigmamap is offset from the center. \\ 
\noindent {\bf J033244.20-274733.5:} We find no evidence for rotation
in the VF. The \sigmamap shows a single peak close to the center of
galaxy, but the central velocity dispersion is much higher than
expected for a rotating system.  \\ 
\noindent {\bf J033250.24-274538.9:} This is a low surface brightness
galaxy with complex kinematics showing no evidence for
rotation. Its large large amplitude of the VF is unexpected for
such a low inclination system. Both VF and \sigmamap show deviations
from a normal RD. \\

\section{Discussion}
\label{secdis}

{\scriptsize
\begin{table*}
\caption{Statistics of each kinematical class in different fields for the complete sample (63 galaxies with $M_J$(AB)$<$$-20.3$).}
\begin{center}
%{\centering
\begin{tabular}{c|ccc|c|c}\hline\hline
   &  HDFS  & CFRS22h   & CFRS03h & CDFS & Total ( fraction) \\ \hline
RD &  4  & 2 & 5 & 9  & 20 (32\%$\pm$12\%) \\
PR &  2  & 2 & 2 & 10 & 16 (25\%$\pm$12\%) \\   %% notice, minus one PR for M_J mag.
CK &  3  & 2 & 6 & 16 & 27 (43\%$\pm$12\%) \\ \hline
UC & \multicolumn{3}{c|}{3 in total} & 3 & 6 (9\%)  \\  %% NOTE: the 3 gal of CDFS have MJ<-20.3; the 3 gal of GTO
\hline \hline
\end{tabular} 
\end{center}
%}\\
Note: the first column gives the kinematical classifications (see
Sect.~\ref{seccls} for details): RD-rotating disks; PR-perturbed rotations;
CK-complex kinematics; UC-unclassified.  Five galaxies
(CFRS220619, CFRS220919, HDFS4070, HDFS4090, J033243.62-275232.6) are
not included in this statistic because they are fainter than the
limiting magnitude of $M_J{\rm (AB)}\!=\!-20.3$.
\label{tbCV}
\end{table*}
}

Table~\ref{tbCV} shows how the galaxies in our sample fall into the
different kinematical classes. We find 32\% RDs, 25\%
PRs and 43\% CKs which are limited
by an error of 12\%.  This confirms the preliminary result of Flores
et al.\ (2006) that at $z\!=\!0.4\,\textrm{--}\,0.75$, few massive
emission line galaxies are kinematically relaxed. Furthermore, the
combined two samples include galaxies from 4 different fields, namely
the CDFS, HDFS, CFRS03h and CFRS22h. We refer to the
later three fields as the GTO sample (see Flores et al.\ 2006).  It is
based on 3 different lines of sight, therefore its result is unlikely
to be affected by the cosmic variance effect.  The GTO sample gives
RD:PR:CK $=$ 11:6:11 (i.e., 39\%:22\%:39\%), which is consistent with
the results from CDFS (25\%:28\%:47\%) within the statistical error.
Notice however that in the field of CDFS we discovered a relatively
large fraction of CKs, which is possibly related to the presence of
large-scale structure (Ravikumar et al.\ 2007).  In total our sample
includes 6 galaxies that are unclassified (UC), because the data are
not well resolved spatially.  The relatively small fraction of
RD galaxies is intriguing, in particular if compared with
the significantly larger fraction of disk galaxies found at low
redshift (see below). We emphasize the robustness of this result, in
particular because:
% difference between item 1 and item 3?
\begin{itemize}
\item {the sample is representative of galaxies with $M_{\rm stellar} \ge
1.5\times 10^{10}M_{\sun}$ (see Sect.~\ref{secrepsample} and
Fig.~\ref{fighist});} 
\item {it is unaffected by cosmic variance since galaxies are selected
from four independent fields, and we find that the fraction of
galaxies with a particular classification does not vary significantly
from field to field (see Table~\ref{tbCV});}
\item {it is based on a representative sample of 63 galaxies, and the
uncertainties to the above fractions are smaller than 12\%.}
\end{itemize}

Let us now consider the general population of galaxies at \zzz, with
$M_{\rm stellar}\!\ge\!1.5\!\times\!10^{10}M_{\sun}$, including
emission line galaxies (with $W_{0}(\mbox{{\sc [Oii]}})\!\ge\!15$\,\AA)
like those studied in this paper, and also more quiescent galaxies with
very faint or without emission lines (e.g., quiescent late type
galaxies such as E/S0 and some early type spirals). Hammer et al.\
(1997, see also Hammer et al.\ 2005) have found that at $z\!=\!0.65$
(the average redshift of our sample), 60\% of intermediate mass
galaxies have emission lines, and this result has been confirmed by
all galaxy surveys. We further assume that all quiescent galaxies have
relaxed kinematics, e.g., pressure (or dispersion) supported for spheroids and
rotationally supported for disks. This implies that at \zzz, 
at least $41\!\pm\!7$\% of galaxies are not dynamically relaxed,
including $26\!\pm\!7$\% of galaxies with CKs.  This
result is in sharp contrast with the kinematics of present-day
galaxies, which are almost all relaxed, and indicates a strong
evolution over the last 5\,Gyr. Indeed, at $z=0$, we find that 70\% of
intermediate mass galaxies are spirals, while irregulars, compact galaxies and
mergers contribute to less than 1\% (e.g., Hammer et al.\ 2005). 

This result is unlikely to be affected by artefacts of our
methodology. To illustrate this, we will now critically evaluate the possible
sources of error leading to misclassifications, and quantify their impact
on the ensemble properties of our sample. 
As discussed earlier (Sect.~\ref{ssecindividuals}), we suspect that two galaxies 
classified as a perturbed rotation may in fact have an enhanced central
velocity dispersion due to the effect of a prominent star-forming, and
possibly gas-rich, bulge. This does not change our result significantly,
as it would only reduce the fraction of kinematically perturbed galaxies from
41\% to 39\%). Moreover, for objects with small spatial coverage, i.e., galaxies
extending over less than 6
spatial pixels, our classification may be less robust than for more extended
galaxies. This is the case for a few compact galaxies, in particular those
with half light radii less than one GIRAFFE pixel (0.52\arcsec), and
for a few more extended galaxies that have relatively weak emission (i.e.,
a mean S/N $<\,4$). However, these galaxies represent less
than 10\% of the whole sample. For more than 90\% of the sample, the
kinematics are well sampled, with a median spatial coverage of 9 pixels at
S/N $\!>\!4$), allowing us to robustly and uniquely classify
the kinematics. We note explicitly that we verified that the classification
does not depend on the mean S/N of the 
galaxies. Moreover, Flores et al.\ (2006) and Puech et al.\ (2007)
have convincingly shown that galaxies with non-relaxed kinematics are
responsible for the large dispersions of both the Tully-Fisher and the
$j_{\rm disk}$--$V_{\rm max}$ relationships. So it is beyond doubt
that a significant fraction of \zzz\ galaxies have kinematics that
deviate significantly from those of their local descendants, i.e., the
present-day intermediate-mass galaxies, which include 70\% of spirals.

Which physical process could explain such a dramatic
evolution in the kinematics of galaxies within a relatively modest amount of
time (4\,--\,7\,Gyr)? The morphology of several galaxies in the sample
strongly advocates that merging is one such process. For example, a 
minor merger may cause perturbed rotation: the in-fall of a
gas-rich satellite would unavoidably lead to a local increase of the
dispersion shifting the peak of the $\sigma$-map. A major merger will
significantly affect a rotational VF by destroying the pre-existing disk, and
lead to a signature resembling a complex VF (see Puech et al.\ 2006a and
2007). It is then probable that merging may explain most of the
discrepancies in the observed VFs at \zzz. However, this fact alone
does not necessarily imply that merging is the only physical process
explaining the strong evolution of galaxy kinematics.

If major mergers are responsible for complex VFs, then 
26$\,\pm\,$7\% of the galaxies within \zzz\ will either be on-going
mergers or merger remnants. This has to be compared with only
5$\,\pm\,$1\% of on-going mergers, as revealed by pair counts, two-point
correlation or morphological classifiers (see a summary in Hammer et
al.\ 2007 and references therein). Indeed, these morphological analyzes
essentially account for the approaching phase of a merger, while
kinematics are affected by large scale peculiar motions induced before and
after the merger. This leads Hammer et al.\ (2007) to postulate that
the merger remnant phase may be 4 to 5 times longer than the
approaching phase. Assuming 0.35\,--\,0.4\,Gyr for the duration of the
latter, this results in a merger remnant phase with a duration of
1.5\,--\,2\,Gyrs. Indeed, simulations by Robertson et al.\ (2006) and
Governato et al.\ (2007) predict such a duration for the rebuilding of
a disk after a major merger of gas rich galaxies.  Furthermore, the requirement for gas rich interacting galaxies in Robertson et al.\ (2006),
finds some support from the evolution of the gas content in galaxies as a
function of cosmic time, although this evolution is derived indirectly from 
the gas phase metal abundance in distant galaxies. Liang et al.\ (2006)
estimated that the gas content in intermediate-mass galaxies at
$z\!\sim\!0.6$ was two times larger than in galaxies at the current
epoch. 

However, the predominance of mergers, and especially major mergers,
leads to the requirement that many present-day galactic disks have
in fact been rebuilt at a relatively recent epoch. Taking into account the 
number fraction of both on-going mergers and galaxies with complex
VFs, Hammer et al.\ (2005, 2007) conclude that between 50\% and 70\% of
galaxies may have experienced a major merger and subsequent disk
rebuilding since $z\!=\!1$. Although this alternative may explain many
observations in the $z\!=\!0.4\!-\!1$ redshift range (e.g., Hammer et
al.\ 2005), could a less dramatic mechanism be at the origin of the
peculiar kinematics \zzz?

In fact our observations can only account for the large scale
motions traced by the ionized gas and not those of the stellar
component. Observations of the latter are certainly crucial, although
for most distant galaxies this is beyond the reach of 8 to 10 meter
telescopes. One may then postulate that rapid gas motions may be
superimposed on a normal rotational stellar component. There are two
difficulties with such an assumption. First, at \zzz\ the gaseous
fraction is much higher than today, and represents a significant
fraction of the baryonic mass. To illustrate this, recall that since
$z\,=\,1$, about half of the present-day stellar mass has been formed
from gas in intermediate mass galaxies (Hammer et al.\ 2005; Bell et
al.\ 2005). One may then wonder how under such conditions an ordered
rotational VF of about half of the baryons (stars) may survive against
highly perturbed VFs for the other half of the baryons (gas). The origin
of a large-scale motion gas component is certainly a second
difficulty, especially if no merger is advocated. Another possibility
may be to invoke internal processes, such as bars,
that may perturb the gaseous VFs. One may wonder whether the presence of bars in the central
region could create additional dispersion leading to apparently 
unrelaxed VFs according to our VF diagnostic (see
Fig.~\ref{figedmod}). However, our spatial resolution may be too
poor to kinematically characterize most of the bars, except possibly the most giant ones. Could some vigorously enhanced internal processes contribute to the complex VFs? Indeed, this is not expected from the apparent non-evolution of the frequency of barred
galaxies (e.g., Zheng et al.\ 2005 and references therein), which does not
support that bars are the process to explain the large evolution of
galaxy kinematics. The observed motions at large scales rather suggest another mechanism, probably related to asymmetric gas accretions (such as provided by a merger) or perhaps to gas outflows.

\section{Conclusion}
We have been able to measure the VFs of 36 galaxies at
\zzz\ using deep exposures of the multiplex integral-field spectrograph
GIRAFFE at the VLT in the multi-IFU mode, measuring the kinematics of the
spatially and spectrally well resolved \oii\ emission line doublet. In
combination with a similar study by Flores et al.\ (2006), we have a 
relatively large and representative sample of 63 galaxies
with $M_{\rm stellar} \ge 1.5\times 10^{10}M_{\sun}$. Thus, our results are
representative for the population of intermediate-mass galaxies in this
redshift range, and it cannot be
affected by cosmic variance. To date, this is the only existing
representative sample of distant galaxies with measured VFs.

We confirm and consolidate the results of Flores et al.\ (2006),
that a significant fraction of intermediate mass galaxies had
perturbed or complex kinematics 5 Gyrs ago. Our method to classify the
kinematics of the galaxies is particularly robust. It attributes a large
weight to the velocity dispersion in the central region of the galaxy, where
the S/N are the highest. Even if we assume that all 
quiescent galaxies at \zzz\ had well ordered VFs, we find
that 41$\,\pm\,$7\% of the galaxies are not kinematically relaxed,
including 26$\,\pm\,$7\% of galaxies that show complex kinematics.
Undoubtedly, galaxy kinematics are evolving very rapidly, since most
present-day galaxies in the same mass range are likely to have ordered
VFs.

This result may be combined with the fact that anomalous VFs
are responsible for most of the large observed dispersion of
both the Tully-Fisher and the $j_{\rm disk}$--$V_{\rm max}$
relationships (see Flores et al.\ 2006 and Puech et al.\ 2007). It
suggests a random walk evolution of galaxies related to a high fraction
of merging events, including major mergers (Puech et al.\ 2007).
Mergers may indeed reproduce all the peculiar kinematics at \zzz, as
well as being responsible for the dispersion of fundamental relations.
Other mechanisms, such as in-fall of high velocity gas clouds, gas outflows or bars may
also contribute to the observed evolution in the kinematics. To understand their influences, and moreover the underlying mechanisms that activate them, requires detail analyzes of individual objects and as a prerequisite, a full model of the significance of the GIRAFFE measurements. If major merging is the
main mechanism responsible for the large fraction of complex VFs, 
then this implies that, since $z\!=\!1$, from 50\% to 70\% of
intermediate mass galaxies have experienced a major merger. This
is quantitatively in good agreement with the spiral rebuilding scenario as proposed by Hammer et al.\
(2005).

\begin{acknowledgements}
We would like to thank the referee for helpful comments and discussions on the paper.
We thank all the GIRAFFE team at the Observatories in Paris and
Geneva, and ESO for this unique instrument. We are grateful to Albrecht
R{\"u}diger for helping us in the writing of the paper. We thank Francoise Combes and Denis Burgarella
for their useful comments and for their kind support of this program since the beginning, as well as
Andrea Cimatti, Emanuele Daddi, David Elbaz, Olivia Garrido, Dominique Proust, Xianzhong Zheng.
\end{acknowledgements}

\clearpage

%\clearpage
%\newpage

\end{document}